\documentclass[aps,prb]{revtex4-1}  
\usepackage[version=3]{mhchem}	
\usepackage{graphicx}  
\usepackage{nicefrac}  
\usepackage{siunitx} 	
\usepackage{amsmath,amssymb,amsfonts}   
\usepackage{hyperref}
\usepackage{import}
\usepackage{mathdesign}
\usepackage{bbold}
\usepackage{wasysym}
\usepackage{xtab}
\usepackage{xtab,afterpage}
\usepackage{lipsum}
\usepackage{multirow}
\usepackage{bm}

\usepackage{isomath}
\usepackage{bbm}
\usepackage{braket}

\newcommand{\hmn}[1]{
  \ensuremath{\begingroup\setupHMN #1\endgroup}%
}

\newcommand{\setupHMN}{%
  \doHMN{-}{\HMNoverline}%
  \doHMN{*}{\HMNminverse}%
  \doHMN{`}{\HMNminversep}%
  \doHMN{~}{\HMNminversea}%
  \doHMN{!}{\HMNminversemtilde}%
  \doHMN{i}{\infty}
}

\newcommand{\doHMN}[2]{%
  \begingroup\lccode`~=`#1
  \lowercase{\endgroup\let~}#2%
  \mathcode`#1="8000
}

\newcommand{\HMNminverse}[1]{\frac{#1}{m}}
\newcommand{\HMNminversep}[1]{\frac{#1}{m'}}
\newcommand{\HMNminversea}[1]{\frac{#1}{a}}
\newcommand{\HMNminversemtilde}[1]{\frac{#1}{\tilde{m}}}
\newcommand{\HMNoverline}[1]{\mkern1mu\overline{\mkern-1mu#1\mkern-1mu}\mkern1mu}

\newcommand{\mat}{\matrixsym}
\newcommand{\ten}{\tensorsym}

\hypersetup{colorlinks=true,urlcolor=blue,citecolor=blue,linkcolor=blue,breaklinks}
\renewcommand{\vec}[1]{\mathbf{#1}}

\renewcommand{\arraystretch}{0.75}

\begin{document}

\title{A tensorial approach to `altermagnetism'}

\author{Paolo G. Radaelli}
\affiliation{Clarendon Laboratory, Department of Physics, University of Oxford, Oxford, OX1 3PU, United Kingdom}
\email[Corresponding author: ]{p.g.radaelli@physics.ox.ac.uk}

\date{\today}

\begin{abstract}
I present a tensorial approach to the description of $\vec{k}/-\vec{k}$-symmetric, time-reversal-odd splitting of electronic bands in magnetic materials, which can be of non-relativistic origin and was recently given the name of `altermagnetism'.  I demonstrate that tensors provide a general framework to discuss magnetic symmetry using both spin groups and magnetic point groups, which have often been contrasted in recent literature.  I also provide a natural classification of altermagnets in terms of the lowest-order tensorial forms that are permitted in each of the 69 altermagnetic point groups.  This approach clarifies the connection between altermagnetism and well-known bulk properties, establishing that the vast majority of altermagnetic materials must also be piezomagnetic and MOKE-active, and provides a rational criterion to search for potential altermagnets among known materials and to test them when the magnetic structure is unknown or ambiguous.  \end{abstract}

\maketitle

\section{Introduction}

In the past two decades, there has been a resurgence of interest in compounds having electronic bands with lifted spin degeneracy, partly motivated by the requirement of new materials for spintronics.  In addition to spin polarisation in ferromagnets, it is well known that spin degeneracy can be lifted even in non-magnetic materials by the famous Rashba-Dresselhaus (R-D) effect,\cite{Dresselhaus1955, Rashba1959} which requires spin-orbit coupling (SOC) and is therefore largest in the presence of heavy elements.  The R-D effect also requires the absence of inversion symmetry, \cite{Manchon2015, Bihlmayer2015} due either to the bulk crystal structure being acentric or to symmetry breaking at interfaces.

Starting from 2019, several groups came to the surprising realisation that spin degeneracy can also be lifted in some fully compensated  antiferromagnets (AFM) (including \emph{collinear} AFM), due to the interaction between electron spins and the `effective Zeeman field' (largely of magnetic exchange origin) produced by ordered magnetic moments.  This effect is clearly distinct from the R-D effect (most notably, it is $\vec{k}/-\vec{k}$ symmetric) and, crucially, does not require SOC, opening the possibility to observe large spin splitting even in light-element compounds.  In these cases, the topology of the associated electronic structures gives rise to macroscopic properties, such as the anomalous Hall effect, that are usually associated with ferromagnets. \v{S}mejkal \textit{et al.}  were among the first to realise this in the context of the spontaneous Hall effect.\cite{Smejkal2020}  \v{S}mejkal \textit{et al.}\cite{Smejkal2020} and Ahn \textit{et al.} \cite{Ahn2019} independently discussed the case of RuO$_2$, which was later to become the `poster child' for this emerging research field.  More or less at the same time, Naka \textit{et al.} described how spin currents can be generated in an organic collinear AFM ($\kappa$-Cl) due to a very similar underlying mechanism.\cite{Naka2019, Naka2020}  Later, the same group focussed specifically on spin splitting in momentum space, which is interpreted as arising from ordering of microscopic multipoles; \cite{Hayami2019, Hayami2020, Hayami2024} they also demonstrated that spin current generation, promoted by site-dependent anisotropic electronic transfer, can occur in simple inorganic compounds such as certain perovskites.\cite{Naka2021}.  Using Density Functional Theory (DFT), Yuan \textit{et al.} demonstrated spin splitting at `atomic-like energy scales' in the light-element insulator MnF$_2$,\cite{Yuan2020} which is a fully compensated collinear AFM.  Although insulators provide excellent proofs of principle, applications of AFM-induced spin splitting for spintronics would be greatly facilitated by the discovery of metallic systems displaying this effect.  The same group later proposed a rational criterion to search for such materials among both collinear and non-collinear AFM, and investigated the electronic structure of several candidates by DFT. \cite{Yuan2021}  Building on their earlier work and using DFT calculations, \v{S}mejkal \textit{et al.}  demonstrated large spin splitting in the absence of SOC in several metallic/semiconducting candidates, including RuO$_2$, CrSb and MnTe,\cite{Gonzalez-Hernandez2021,Smejkal2022, Smejkal2022b}.  These authors pioneered the use of spin group analysis (see below) to define and delimit the relevant phenomenology in collinear AFM, and coined the term `altermagnetism' to mark the distinction between non-spin-split AFM, ordinary spin-split ferromagnets with uniform uncompensated  spin splitting and these materials, which display \emph{alternating} compensated spin splitting.  

This terminology has been widely used in subsequent literature, which includes both theoretical\cite{Liu2023, McClarty2024, Cheong2024} and experimental\cite{Samanta2020, Feng2022, Bai2023,Reimers2024,Osumi2024, Krempasky2024} contributions. 
However, throughout the literature, different authors employ somewhat different definitions of altermagnetism. \footnote{ For example, Cheong \textit{et. al.} \cite{Cheong2024} define `Type-III altermagnets' as systems with symmetries including time reversal is a distinct operator.  For reasons explained at length in this paper, this can only give rise to  $\vec{k}/-\vec{k}$ \emph{anti}symmetric textures of the R-D type, though magnetism can play a very important role in them.  Here, I am not considering Type-III altermagnetism and, more generally magnetism-induced R-D type textures, though I am including a brief discussion of these effects in sec. \ref{sec: discussion} .}  \v{S}mejkal's emphasis was very much on \emph{collinear} metallic antiferromagnets, for which spin-splitting effects should be stronger, DFT is much easier to perform, and which may have advantages for certain spintronic applications.  However, Yuan \textit{et al.}\cite{Yuan2021} and, more recently, Cheong and Huang \cite{Cheong2024} showed that altermagnetic-like spin textures can also be expected in many non-collinear magnets, which may present distinct practical advantages and should not be excluded \textit{a priori}.   

From the very beginning, symmetry analysis has been recognised as an essential tool of this research fields, since the guiding principle to select potentially altermagnetic materials must lie in their underlying symmetry properties.  Yuan \textit{et al.}  \cite{Yuan2020} proposed an approach based on magnetic space groups (MSG), while  \v{S}mejkal \textit{et al.} proposed the so-called spin groups (SGs) \cite{Litvin1977, Smejkal2022, Liu2022}, which allow for extra (approximate) symmetries to be considered when spins and the lattice are nearly decoupled (see below).  Cheong propose a classification based on magnetic point groups (MPGs) \cite{Cheong2024}, while Liu \textit{et al.}\cite{Liu2023} employ little co-groups at specific points of the Brillouin zone (BZ), and propose a classification that also includes higher-order effects (quartic altermagnetism). 


The interest in altermagnetism in non-collinear magnetic structures\cite{Cheong2024} and in `weak-altermagnetic' effects \cite{Krempasky2024} undoubtedly requires an extension of the SG approach as employed in ref. \onlinecite{Smejkal2022}. \footnote{Ref. \onlinecite{Smejkal2022} employs `binary' SGs, i.e., SG with 2-element groups acting on spins (see below).  Binary SG  cannot be used for non-collinear structures.  Certain non-collinear structures can be described either with more complex SGs \cite{Litvin1977}  or with so-called multi-colour groups,\cite{Harker1981} and in the framework of the exchange multiplet theory \cite{Izyumov1980}.  To my knowledge, a complete theory that parallels that of binary SGs has not been developed thus far for non-collinear structures.  See Appendix \ref{App: spin_groups} for more details.} MPGs are a flexible tool to deal with these cases, especially since there is no reason to expect that altermagnetic splitting should always be weak for non-collinear structures.  Intuitively, a classification based on MPGs should be entirely adequate, since spin splitting ultimately results in anomalous \emph{macroscopic} properties (e.g., the anomalous spin Hall effect) that are subjected to MPG symmetry constraints via the Neumann's principle.\footnote{The Neumann-Minnigerode-Curie Principle (NMC Principle) enables one to derive the selection rules for the physical properties from the symmetry of the crystal (or molecule) in question. For macroscopic properties, the relevant symmetry is precisely the MPG of the crystal.  One should also remark that, unlike the widely used MPGs, magnetic \emph{space} groups have been almost entirely superseded among magnetic structures specialists by irreducible representations analysis `\`a la Bertaut',\cite{Bertaut1968} with additional symmetries in spin spaces being dealt with using Izyumov's exchange multiplet theory \cite{Izyumov1980}} However, one should not lose sight of the advantages of SGs when it comes to pseudo-symmetric structures, i.e., when the crystal symmetry is only `slightly' broken upon magnetic ordering.\footnote{Perhaps the best example of this is ferromagnetic ordering in cubic magnetic metals such as Fe or Ni.  Strictly speaking, Fe becomes rhombohedral below the Curie temperature, but the deviation from cubic symmetry due to magneto-striction is extremely small ($\approx 10^{-6}$).}  It seems fair to state that, in previous work, the link between MPGs, SGs and the complexity of spin textures in reciprocal space has not been entirely clarified.

Here, I propose an approach based the complete expansion of  altermagnetic spin textures in momentum space in terms of Cartesian and spherical tensors.  This approach can be equally applied to vectorial textures (described by MPGs) and scalar textures (described by binary SGs, referred to simply as as SGs in the remainder), though my emphasis will initially be on the former.  Starting from a working definition of `altermagnetic textures' as being simply those that are $\vec{k}/-\vec{k}$-symmetric and time reversal odd, I demonstrate that, to any given order, the expansion of  vectorial altermagnetic textures is described by a single Cartesian tensor of odd rank, while R-D ($\vec{k}/-\vec{k}$-antisymmetric, time-reversal even) vectorial magnetic textures are described by even-ranked tensors.  I also propose a natural classification of such textures based on the lowest-order tensor forms allowed in a particular set of MPGs (class), and show that the vast majority of altermagnetic point groups  (66 out of 69) display quadratic (rank-3) altermagnetism, with the remaining three groups allowing quartic (rank-5) altermagnetism.  For the case of collinear AFM structures, I demonstrate that, in most cases, SG scalar textures correspond to the component of the MPG vector magnetic textures along the direction of the staggered magnetisation, while the MPG approach also produces a pattern of `weak altermagnetic' textures in other directions.  One notable exception is represented by cubic groups, since the pseudo-cubic symmetry introduces constrains between texture components that would otherwise be independent.  For these cubic groups, I provide a conversion table between SG and MPG textures for various direction of the magnetic moments, so that one could remain entirely within the MPG framework, generally the more familiar to the magnetism community.  Performing the MPG and SG analyses \emph{in parallel} has the distinct advantage of clearly separating the components of the textures that do and do not depend on the direction of the staggered magnetisation, which generally have very different energy scales.

One important practical implication of the tensor approach is the connection with macroscopic properties:  materials displaying \emph{quadratic} altermagnetism (66 out of 69 MPGs) must also allow the piezomagnetic effect and the magneto-optical Kerr effect (MOKE)  --- two easily-testable phenomena that have been known for several decades \cite{Dzialoshinskii1957, Kahn1969}, and for which extensive materials databases are available (for example, ref. \onlinecite{Gallego2016}).  Testing for the presence of these effects might also facilitate the screening of candidate altermagnetic materials before more complex experiments are performed to to measure AFM-induced spin splitting of electronic bands directly.

This paper is organised as follows: in secs. \ref{sec: Tensorial description}, \ref{sec: conditions} and \ref{sec: Jahn}, I present a general tensorial treatment of spin textures in momentum space, which can be applied to both altermagnetic and R-D-type textures.   In sec. \ref{sec: Jahn}, I also outline how a parallel treatment for can be performed for \emph{vector} and \emph{scalar} textures, described by MPGs and binary SGs respectively, and show that dealing with the latter is greatly facilitated if one employs time reversal (via Shubnikov groups) instead of 2-fold rotations.  In secs. \ref{sec: tensor_forms}, \ref{sec: sperical_decomp} and \ref{sec: hexadec}, I establish which altermagnetic tensor forms are allowed by symmetry and propose a classification of MPGs based on the lowest-rank altermagnetic tensors.  Additional symmetry constraints at special points in reciprocal space are discusses in sec. \ref{sec: sym_rec_sp}.  Up to this point, no distinction is made between strong and weak effects, in keeping with the classical Neumann approach.  In secs. \ref{sec: collinearity} and \ref{sec: higher_orders}, I discuss the special case of collinear structures and explain how the dominant components of the spin textures, which are expected to be parallel to the N\'eel vector, can be extracted from the tensors.  In sec. \ref{sec: spin_group_cubic}, I explain the necessity of employing SGs (which I do via tensorial analysis) for systems that are crystallographically cubic and in which symmetry is only broken upon magnetic ordering, whilst also discussing the application of spin groups to cases in which the magnetic symmetry is very low.  Numerous examples in different symmetries are presented in sec. \ref{sec: examples}, including graphical depictions of several types of spin textures.  The paper is concluded by a summary and general discussion (sec. \ref{sec: discussion}).


\section{The tensorial description }
\label{sec: Tensorial description}

The starting point to develop a tensorial description of altermagnetism and, more generally, of spin splitting of electronic bands, is the expansion in Cartesian tensors of the reciprocal-space spin textures, i.e., the vector field of spin states in momentum space.  This semiclassical vector field, which I will denote as $\vec{B}^{eff} (\vec{k},m)$, determines both the direction of the spin quantisation axis and the strength of the spin splitting/polarisation, and depends both on the wavevector $\vec{k}$ and on the band index $m$. \footnote{More rigorously, within the context of DFT, for a given wavevector $\vec{k}$ and band index $n$ the spin texture is defined by the vector field  $ \vec{s}_{n\vec{k}}=\bra{\Psi_{n\vec{k}}}\bm{\sigma}\ket{\Psi_{n\vec{k}}}$, where $\bm{\sigma}$ is a vector of Pauli matrices and the integral implied by the $\bra{}$ and $\ket{}$ is over the real-space unit cell.}

\begin{equation} 
\label{eq. tensor_expansion}
B_i^{eff} (\vec{k}, m)=\sum_{l=n}^{n+1} T^{(l)}_{i,\alpha\beta\gamma \dots} k_\alpha k_\beta k_\gamma \dots
\end{equation}
 
The Cartesian tensors $T^{(l)}_{i,\alpha\beta\gamma}$ is of rank $l$ and depends on $k=|\vec{k}|$ and on the band index $m$. In eq. \ref{eq. tensor_expansion}, repeated indices are implicitly summed, and the tensor if fully symmetric over the Greek indices. 
 
The expression in eq. \ref{eq. tensor_expansion} is directly derived from to the expansion of the spin texture into spherical tensors, and is therefore \emph{complete} on each spherical shell in momentum space (or sections thereof near the BZ boundary).  In fact, one can show (see Appendix \ref{App_A1}) that the the sum of two Cartesian tensors of ranks $n$ and $n+1$ is identical to the expansion of the vector field $B_i^{eff}$ onto a spherical basis up to $\Gamma^{1} \otimes \Gamma^{n}$, where $\Gamma^n$ is the irreducible representation (\textit{irrep}) of the group $SO(3)$ with $L=n$ and ``$\otimes$'' is the tensor product of representations.  Note that higher-rank Cartesian tensors need not be necessarily small --- in fact, reproducing the spin texture, especially near the BZ boundary, may require a high-order expansion.  Nevertheless, as I will show in the remainder, a very natural classification can be obtained by considering only low-rank Cartesian tensors.

In the presence of (proper and improper) rotational symmetry combined with time reversal symmetry in an MPG, the tensor forms in eq. \ref{eq. tensor_expansion} are constrained by the requirement that the spin textures have at least the symmetry of the MPG of the crystal, which implies that the tensors themselves must be totally symmetric by all the MPG operations --- in other words, the tensors must transform like the totally symmetric \textit{irrep} of the MPG (see Appendix \ref{App_A2}).

 Hence, for a given tensor rank, tensors at all values of $k$ have the same general form (imposed by symmetry) and their $k$ dependence is described in terms of a reduced number of scalar functions, corresponding to the free parameters of that tensor form.  Additional constraints will occur at the $\Gamma$  point (zone centre), at band crossing points and at special points at the boundary of the BZ with non-trivial little-group symmetry. 
 
When considering inversion (parity) and time reversal symmetries specifically, in the expansion in eq. \ref{eq. tensor_expansion} one can immediately distinguish between \emph{even-rank tensors}, which are time reversal even and parity odd, and \emph{odd-rank tensors}, which are time-reversal odd and parity even.   Once again, it is important to emphasise that  only two Cartesian tensor are required for a truncated spherical tensor expansions, and that all effects allowed by lower-rank tensor are \emph{automatically included} in higher-rank tensors of the same parity.  For example, the rank-2 tensor is at linear order in $\vec{k}$ and corresponds to the usual Rashba-Dresselhaus (R-D) effect, while the rank-4 tensor includes both the linear and cubic R-D tensor etc.  In keeping with recent literature, the rank-3 tensor can be called `quadratic altermagnetic tensor' and includes the ordinary ferromagnetic spin splitting (rank 1), while the next-highest order that is $\vec{k}/-\vec{k}$ symmetric is a rank-5 tensor, which includes both quadratic and quartic altermagnetism as well as ferromagnetic spin splitting. 

A completely parallel treatment to the one I just outlined can be performed for \emph{scalar} parity even and time-reversal-odd textures in real/reciprocal space, which are used to describe collinear magnetic structures within the SG framework.  This becomes completely transparent if one casts binary SGs in the slightly different language of time reversal rather than two-fold rotations (see Appendix \ref{App: spin_groups} for a complete discussion).  The corresponding tensors are totally symmetric and their rank is lowered by one with respect to the corresponding vectorial textures.


\section{Conditions for altermagnetism and the Rashba-Dresselhaus effect}
\label{sec: conditions}

Altermagnetism thus defined and for all odd ranks requires the following minimum requirements to be satisfied:

\begin{enumerate}
\item Time reversal symmetry must be broken --- in other words, the effect is only allowed in ferromagnets (FM)  and some antiferromagnets (AFM).
\item For a given point $\vec{k}$ in the BZ, the spin splitting must reverse in the time-reversed FM or AFM domain.
\item Spin splitting must be \emph{symmetric} between $\vec{k}$ and $-\vec{k}$
\end{enumerate}

These conditions were thoroughly discussed in ref. \onlinecite{Yuan2020}, and exactly mirror those for the R-D effect at any even-rank tensor:

\begin{enumerate}
\item Inversion symmetry must be broken.
\item If the material is magnetic, for a given point $\vec{k}$ in the BZ, the spin splitting must be the same in the time-reversed FM or AFM domains.
\item Spin splitting must be \emph{anti-symmetric} between $\vec{k}$ and $-\vec{k}$
\end{enumerate}

Note that spin splitting is not possible at any order in the presence of $PT$ symmetry (i.e., if the product of inversion and time reversal, $\theta I$, is a symmetry operator)--- a well-known result that can be obtained from simple symmetry considerations. \cite{Tang2016}  Nevertheless, altermagnetism is \emph{not} incompatible with the magneto-electric effect, tough the latter is often associated with $PT$ symmetry conservation.  \footnote{An example of this is MPG $3m'$, which is admissible (i.e., compatible with ferromagnetism), polar, magneto-electric and altermagnetic. }

\section{Jahn symbols and the connection with macroscopic properties}
\label{sec: Jahn}
 
Further progress can be made by defining the so-called Jahn symbol for these tensors,\cite{Jahn1949} which for a tensor of rank $n$ in eq. \ref{eq. tensor_expansion} is generically:
 
\begin{equation} 
\label{eq: generic_Jahn}
ea^nV[V^{n-1}]
\end{equation}
 
where $e$ indicates that all these are pseudo-tensors (with opposite parity to that of ordinary tensors, which are parity-even for even ranks and parity-odd for odd ranks), $a$ defines the time reversal properties ($a^n$ is time-reversal-even when $n$ is even, odd when $n$ is odd) and the square brackets indicate symmetrisation on all indices, since these are contracted with the indices of $\vec{k}$.

The advantage of this notation is that one can sometimes connect these tensors with tensors defining apparently unrelated \emph{macroscopic} properties that have the same Jahn symbol.  One can then mine the often extensive materials databases for these known effects and extract candidate materials where the `novel' effect is allowed by symmetry. For example, The Jahn symbol for the linear R-D effect is $eV^2$, which is the same as the magnetotoroidic tensor, while the Jahn symbol for the cubic R-D effect is $eV[V^3]$, which does not have an obvious macroscopic counterpart.  Most relevant for this paper, the quadratic altermagnetic tensor $eaV[V^2]$ has the same Jahn symbol as the tensors describing the piezomagnetic effect and the MOKE effect. Piezomagnetism has been known since the 1950s, \cite{Dzialoshinskii1957, Borovik-Romanov1960}  and a symmetry classification can be found in the famous book by Robert Birss \cite{Birss1966}, first published in 1964, which also lists several materials now discussed in the context of altermagnetism.  Interest in MOKE activity in antiferromagnets and weak ferromagnets is almost as old, dating back to the 1960's, \cite{Kahn1969, Smolenskii1975,Zubov1988,Zenkov1989, Eremenko1992}, and has recently experienced a resurgence. \cite{Higo2018,Kang2022}  

For \emph{scalar} altermagnetic textures, the corresponding Jahn symbols to those in eq. \ref{eq: generic_Jahn} are $a[V^{n}]$, where $n$ is even.  So, for example, the tensor for scalar textures corresponding to quadratic altermagnetism has Jahn symbol  $a[V^{2}]$, one rank lower compared to the corresponding vectorial texture $eaV[V^2]$ but with the same parity/time reversal properties.


 \section{Tensor forms for quadratic altermagnetism} 
 \label{sec: tensor_forms}
 
The allowed altermagnetic tensor forms are restricted by the MPG symmetry of the crystal, so that the tensor itself is completely invariant by any of the MPG operations.  The number of free parameters in each tensor form corresponds to the number of times the totally symmetric \textit{irrep} is contained in the tensor representation.   In general a particular tensor form is shared by several MPGs, which can thus be grouped into `classes'.   For each  MPG, the determination of the number of free parameters via character decomposition and the construction of the appropriate tensor form by projection is greatly facilitated by employing the MTENSOR tool provided by the Bilbao Crystallographic Server. \cite{Gallego2019} Tensors up to rank 6 were calculated using MTENSOR.  For tensors above rank 6, polynomial forms were obtained by the standard projection (symmetrisation) method (see for example sec. \ref{sec: cubic_groups}).

Table \ref{Table_1} lists the 17 unique symmetry-adapted quadratic altermagnetic tensor forms, together with the set of MPGs  (class) that share that tensor form, the number of free parameters and a simplified spherical tensor decomposition (further discussed in section \ref{sec: sperical_decomp}).\footnote{\label{foot: conventions} I employed the conventions adopted by the program MTENSOR of the Bilbao Crystallographic Server (which are somewhat different from those in ref. \onlinecite{Opechowski1965}), except for $\bar{6}'m'2$, which has been converted to  $\bar{6}'2m'$ to be included in Class XIV.  Note that the textures of Classes X and XI differ by a $45^\circ$ rotation, and are therefore equivalent at the MPG level, only distinguished  by the orientation of the texture with respect to the crystal axes --- see further discussion in section \ref{sec: examples}.  In the interest of practical use and to follow conventions, I kept these two classes separate.  Straightforward axes transformations may be required for particular MPGs}  I employed the usual convention, which applies for example to the piezoelectric tensor, in which a $3 \times 6$ matrix is contracted with the array

\begin{equation}
\left[k_x^2, k_y^2, k_z^2, k_y k_z,  k_x k_z, k_x k_y\right]  
\end{equation}


to yield the three-component $\vec{B}^{eff}$.  The effective field $\vec{B}^{eff}$ depends on one or more scalar functions $\Lambda_{ij} (k)$ that are constant on each spherical shell in momentum space (for a note on axes conventions, see~  \footnote{I adopted the standard conventions for $x$, $y$, and $z$, which are related to the direction of the symmetry directions.  For example, for point group $3m1$, $z$ is parallel to the 3-fold axis, $x$ is perpendicular to the mirror plane and $y$ completes the set.  In the setting $31m$ of the same point group, $x$ and $y$ are interchanged.  Examples on how to deal with conventions issues are provided in section \ref{sec: examples}.}).  Examples of this procedure and on how to impose constraints at special points of the BZ are provided in section \ref{sec: examples}.

Table \ref{Table_1} also includes the single class of altermagnetic MPGs that do not allow quadratic altermagnetism (Class XVIII), together with the expression for their spin texture, described by a rank-5 spherical magnetic hexadecapole (see discussion in section \ref{sec: hexadec}).

Among the 17 quadratic altermagnetic classes, eight classes (marked in \textbf{bold} and with a (\S) in Table \ref{Table_1}) allow a magnetic dipolar component, i.e., they allow a net magnetic moment and uncompensated spin splitting, though spin polarisation need not arise primarily from and be parallel to the net moment.  The 31 magnetic groups in these classes are defined in ref. \onlinecite{Cheong2024} as `Type-I altermagnetic', and  coincide with the \emph{admissible} point groups, i.e., the ones allowing a net ferromagnetic moment.  Out of the remaining ten classes, \footnote{These classes are named `Type-II altermagnetic' in ref. \onlinecite{Cheong2024}.} six classes (VIII, XII, XIV, XV, XVII and XVIII, marked in Table \ref{Table_1} in \textit{italic} and with an asterisk $^*$) do not allow the spin quantisation axis to be along an allowed collinear antiferromagnetic direction at rank-3 tensor level, while the other four classes (V, IX, X and XI, marked in Table \ref{Table_1} with a dagger $\dagger$), do allow it.  This aspect will be discussed further in section \ref{sec: collinearity}.

\section{Spherical tensor decomposition}
\label{sec: sperical_decomp}

In addition to the Cartesian tensor form, it is often useful to decompose tensors onto a symmetry-adapted spherical basis, which makes the symmetry constraints more transparent.  Moreover, the spherical (multipole) decomposition creates a bridge between MPG-based approaches and theories based on multipoles, such as the Cluster Multipole Theory \cite{Suzuki2017, Hayami2019, Hayami2020} and the Landau approach to ferro-multipolar ordering proposed by P. McClarty.\cite{McClarty2024}$^,$ \footnote{The connection with the latter is most obvious, because ferro-ordering of a given multipolar form is allowed if and only if this form is totally symmetric by the MPG operations.  However, I stress that here the symmetry-adapted spherical tensor forms were obtained by direct projection onto the totally symmetric \textit{irrep} of each MPG, rather than by analysing the effect of each symmetry operators as it is done in ref. \onlinecite{Suzuki2017}.}

As an example, I recall that the rank-2 R-D tensor can be decomposed into a pseudoscalar ($L=0$), an ordinary polar vector ($L=1$) and a pseudo-quadrupolar traceless tensor ($L=2$), all these being time-reversal even.   Consequently, the linear R-D effect is allowed in \emph{chiral} point groups (those that allow a pseudoscalar), \emph{polar} point groups (which allow an ordinary vector) and a few other non-centrosymmetric groups that only allow the pseudo-quadrupolar traceless tensor (e.g., $\hmn{-4m2}$).  When either of these is allowed, these Cartesian tensors can be written as a linear combination of symmetry-adapted spherical basis tensors.

In the case of quadratic altermagnetism, one can perform an entirely analogous decomposition: let $\Gamma^L$ be the SO(3) \textit{irrep} with dimensionality $2L+1$.  Then

\begin{eqnarray}
\left[V^2\right]&=&\Gamma^0+\Gamma^2\nonumber\\
V\left[V^2\right]&=&\Gamma^1 \otimes \left(\Gamma^0+\Gamma^2\right)=2 \Gamma^1 + \Gamma^2 + \Gamma^3
\end{eqnarray}

Taking into account parity and time reversal, one concludes that the altermagnetic tensor  $ea V\left[V^2\right]$ decomposes into two magnetic dipoles, a magnetic quadrupole and a magnetic octupole.  Since the dipole \textit{irrep} occurs twice, there is some arbitrariness in the decompsition of the dipolar field, but the natural choice is for one of the two components to be \emph{parallel to the dipole vector} ($\ten{D}^I\!$, which is also allowed at the lowest rank-1 order and represent the ordinary ferromagnetic uncompensated spin splitting), while the other is \emph{parallel to the wavevector} $\vec{k}$ ($\ten{D}^{I\!\!I}$).  The quadrupole terms yields a $\vec{B}^{eff}$ that is \emph{perpendicular} to  $\vec{k}$, while  octupolar terms always connect different directions of $\vec{B}^{eff}$ .

With this choice and in the absence of any symmetry we can define the four Cartesian tensors as:

\begin{eqnarray}
\ten{D}^{I}&=&\delta_{ij} \delta_{\alpha\beta} v_j\nonumber\\
&=&
\left(
\begin{array}{cccccc}
 v_x & v_x & v_x & 0 & 0 & 0 \\
 v_y & v_y & v_y & 0 & 0 & 0 \\
 v_z & v_z & v_z & 0 & 0 & 0 \\
\end{array}
\right)
\end{eqnarray}

\begin{eqnarray}
\ten{D}^{I\!\!I}&=&\frac{1}{2}\delta_{i \alpha}\delta_{\beta j} u_j\nonumber\\
&=&\left(
\begin{array}{cccccc}
 u_x & 0 & 0 & 0 & u_z & u_y \\
 0 & u_y & 0 & u_z & 0 & u_x \\
 0 & 0 & u_z & u_y & u_x & 0 \\
\end{array}
\right)
\end{eqnarray}

where $\vec{v}$ and $\vec{u}$ are the two dipolar components.  The quadrupolar tensor can be constructed from a traceless symmetric matrix $S_{jk}$ as

\begin{widetext}
\begin{eqnarray}
\ten{Q}&=&\frac{1}{2}\left( \epsilon_{i\alpha k}\delta_{j\beta} + \epsilon_{i\beta k}\delta_{j\alpha} \right) S_{jk}\nonumber\\
&=&\left(
\begin{array}{cccccc}
 0 & S_{23} & -S_{23} & -S_{11}-2 S_{22} & -S_{12} & S_{13} \\
 -S_{13} & 0 & S_{13} & S_{12} & 2 S_{11}+S_{22} & -S_{23} \\
 S_{12} & -S_{12} & 0 & -S_{13} & S_{23} & S_{22}-S_{11}\\
\end{array}
\right)
\end{eqnarray}
\end{widetext}

 while the octupolar tensor $\ten{O}$,  defined as a linear combination of $L=3$ tesseral harmonics, is:
 
\begin{widetext}
\begin{equation}
\ten{O}=\left(
\begin{array}{cccccc}
 -O_{122}-O_{133} & O_{122} & O_{133} & O_{123} & 2 O_{311} & 2 O_{211} \\
 O_{211} & -O_{211}-O_{223} & O_{223} & 2 O_{322} & O_{123} & 2 O_{122} \\
 O_{311} & O_{322} & -O_{311}-O_{322} & 2 O_{223} & 2 O_{133} & O_{123} \\
\end{array}
\right)
\end{equation}
\end{widetext}

In Appendix \ref{App_B}, Table \ref{Table_2}, a symmetry-adapted spherical basis is presented, comprising nine unique forms (two dipoles, three quadrupoles and four octupoles) plus forms obtained by axis permutation.  In each of the 17 quadratic altermagnetic classes, the unique tensor form can be expressed as a linear combination of one or more symmetry-adapted spherical basis tensors, as shown in Table \ref{Table_3}.  

\section{Quartic altermagnetism}
\label{sec: hexadec}

Having classified all the MPGs that allow rank-3 (quadratic) altermagnetism, one may wonder whether any higher-order altermagnetic point group has been left out --- in other words, are there symmetries where piezomagnetism and MOKE activity are forbidden but higher-order altermagnetism is allowed?  Having excluded all paramagnetic point groups (where altermagnetism is not allowed at any order) and groups that preserve $PT$ symmetry (where spin splitting is entirely forbidden), only three MPGs are left:  $\hmn{432}$, $\hmn{-43m}$, and  $\hmn{m-3m}$, which have been included in Table \ref{Table_1} as Class XVIII.  All these MPGs allow quartic (but not quadratic) altermagnetism and share the same one-parameter rank-5 tensor form, such that:

\begin{equation}
B^{eff}_i=\Lambda (k) \epsilon_{ijl}(k_j^3k_l-k_l^3k_j)
\end{equation}

$ \epsilon_{ijl}$ being the totally antisymmetric Levi-Civita symbol.  The corresponding tensor is a pure $L=4$ (magnetic hexadecapole) spherical tensor, since the $L=5$ spherical representation does not contain the totally symmetric \textit{irrep} of these point groups.

Including quartic altermagnetism will also affect the classification of the MPGs, since some of the classes will split on the basis of their rank-5 tensor forms.  In particular, the tetragonal and hexagonal groups in Classes XIII, XV and XVI will split because they have different rank-5 tensor forms, and Class XVIII will also split into two subclasses (this is indicated with the letters (a) and (b) in Table \ref{Table_1}).  However, \emph{no further splitting will occur at any higher tensor order}, since the MPGs in each class/subclass only differ by the proper/improper nature of some rotations, which has no bearing on any parity-even tensor.  Therefore, the classification of altermagnetic groups in Table \ref{Table_1} can be considered as complete.

\section{Symmetry in reciprocal space}
\label{sec: sym_rec_sp}

Spin textures constructed with the tensorial method I just described have the full symmetry of the MPG of the crystal, which, in turn, means that the field at a given point in the \emph{interior} of the BZ will be locally symmetric by the little co-group of that point, without any need for further symmetrisation. \cite{Liu2023} However, care must be taken at the zone centre and at special points at the BZ boundary, which have additional symmetries due to the fact that points related to them by some symmetry are also related by a reciprocal lattice vector.  Unlike the case of the R-D tensor and more generally of even-ranked tensors, there is no general requirement that the spin texture be zero at Kramers points (i.e., points for which $2 \vec{k}$ is a reciprocal lattice vector).  In general, uncompensated spin splitting is allowed everywhere if the dipolar term $\ten{D}^{I\!}$ is allowed, as is the case for ordinary ferromagnets.  Whenever dipolar terms are \emph{not} allowed, spin splitting must vanish at the zone centre and at zone-boundary points with the same little-group symmetry.  In other cases, a reduced tensor may be obtained by symmetrisation over the index that is not contracted with $\vec{k}$.  Some examples are given in section \ref{sec: examples}.

\section{Collinearity}
\label{sec: collinearity}

Collinearity is not strongly constrained by MPG symmetry, only being forbidden in cubic MPGs.  In all other cases, collinear ferromagnetic or antiferromagnetic structures are allowed provided the following conditions are met:

\begin{enumerate}
\item The magnetic moment is either along the high-order rotation axis (hexagonal, trigonal, tetragonal), along one of the 2-fold axes (orthorhombic) or either parallel of perpendicular to the unique 2-fold axis (monoclinic).  No special condition exists for triclinic groups.
\item The point-group symmetry of the magnetic site is \emph{admissible}, with the admissible direction coinciding with one of the directions specified in point 1. above.  This includes magnetic sites in a general position, since point group $1$ is of course admissible.
\end{enumerate}

It follows that collinear magnetic structures are allowed by symmetry in all altermagnetic groups with the exception of the cubic ones (Classes XVII and XVIII).  However, as anticipated in section \ref{sec: tensor_forms}, one must draw an important distinction between classes that do not allow the spin quantisation axis to be along a permitted collinear direction at the rank-3 tensor level (classes VIII, XII, XIV, XV and of course the purely non-collinear XVII and XVIII, for which I will use the adjective `weak-collinear' ) and those that do (the `strong-collinear' classes V, IX, X and XI).  In the latter case, spin splitting in collinear structures could arise from a very similar effect as for ordinary ferromagnets, in that spins travelling along certain crystallographic directions would experience a net effective Zeeman field originating from the ordered magnetic moments, the quantisation axis switching sign depending on the wavevector direction.  Unlike the case of true ferromagnets, where the magnetisation is uniform, the internal magnetic field distribution in ferrimagnets and antiferromagnets can never be entirely collinear.  Nevertheless, one expects the effective fields (mostly of exchange origin) to be strongest in the direction parallel or antiparallel to the N\'eel vector.  It is noteworthy that all MPGs allowing ferromagnetism can trivially be strong-collinear in the ferromagnetic direction, though some can also be strong-collinear in other directions as well (see example \ref{sec: ex_spingroups}).

By contrast, in the weak-collinear classes (VIII, XII, XIV, XV, XVII and XVIII), the spin quantisation axis produced by the rank-3 tensor is not along one of the allowed collinear antiferromagnetic directions.   However, rank-3 spin splitting in these classes can still be very large for non-collinear magnetic structures.  All weak-collinear classes with the exception of Classes XVII and XVIII (which cannot support collinear structures) become strong-collinear at a higher tensor order --- see sections \ref{sec: higher_orders}.

Examples of a strong-collinear and a weak-collinear class are given in section \ref{sec: examples} with further discussed in section \ref{sec: discussion}.

\section{Strong-collinear higher-order altermagnetism}
\label{sec: higher_orders}

Strong-collinear altermagnetism will appear for all classes (except Classes XVII and XVIII) at some tensor rank, even when it is forbidden by symmetry at the rank-3 (quadratic) level.  Indeed, in their treatment employing SGs, \v{S}mejkal\textit{et al.} \cite{Smejkal2022} list a number of examples that clearly require higher-order tensors, so one needs to show that analogous results are obtained with MPGs.  Classes VIII, XII, XIV and  XVb allow strong-collinear quartic (rank-5) altermagnetism, while Class XVa only allows it at the next order (rank-7).  The expressions for the full tensors are naturally rather complex, but it is easy to write the $\vec{B}^{eff}$ along the collinear direction $z$ (see example in section \ref{sec: Class XV} for more detail).  The appropriate polynomials are reported in Table \ref{Tab: newTab}.  As shown in sec. \ref{sec: examples}, there is a close correspondence between these forms and those obtained by \v{S}mejkal\textit{et al.} using the SG treatment (ref. \onlinecite{Smejkal2022}, fig. 2).  However, it is important to stress that the MPG and SG forms \emph{are not always identical}.  Rather, SG-derived tensor forms are \emph{particular instances} of the corresponding MPG forms, obtained linking the free parameters through particular relationships, which arise from the fact that the SG approach takes into account additional (approximate) symmetries (see. section \ref{sec: spin_group_cubic} for a full discussion). This also demonstrates that the MPG and SG approaches are \emph{mutually consistent} when it comes to strong-collinear altermagnetism.

\afterpage{

\begin{widetext} 


\begin{tiny}
\tablecaption{\label{Table_1} Tensor forms for the 18 altermagnetic classes of MPGs.  Classes I-XVII are the quadratic altermagnetic ($=$ piezomagnetic, MOKE-active) classes, each comprising several MPGs (column 3).   The number of parameters for each tensor form and the spherical tensor decomposition ($D = $ dipole, $Q = $ duadrupole, $O =$ octupole) are provided in columns 4 and 5  (see Appendix \ref{App_B}).  Class XVIII comprises the three remaining altermagnetic groups with rank-5 as the lowest-rank tensor (H $ = $ hexadecapole). For Class XVIII, the explicit form of $\vec{B}^{eff}$ is provided instead of the (very cumbersome) matrix. MPGs labelled with (a) and (b) form separate subclasses when quartic altermagnetism is accounted for (see text).}
\begin{xtabular*}{\textwidth}{l@{\extracolsep{\fill}}lllll}
\shrinkheight{-10pt}

\hline\hline\hline
\multicolumn{5}{c}{\textbf{}}\\
\textbf{Class} & \textbf{Tensor Form} & \textbf{Magnetic point groups} &\textbf{ Parameters} & \textbf{Spherical decomp.}\\
\\
\hline
\\
\textbf{Class I}$^\S$ &
$
\left(
\begin{array}{cccccc}
 \Lambda _{11} & \Lambda _{12} & \Lambda _{13} & \Lambda _{14} & \Lambda _{15} & \Lambda _{16} \\
 \Lambda _{21} & \Lambda _{22} & \Lambda _{23} & \Lambda _{24} & \Lambda _{25} & \Lambda _{26} \\
 \Lambda _{31} & \Lambda _{32} & \Lambda _{33} & \Lambda _{34} & \Lambda _{35} & \Lambda _{36} \\
\end{array}
\right)$ &
\begin{tabular}{cc}
\hmn{1} & \hmn{-1}
\end{tabular} 
& 18 & D(6)+Q(5)+O(7)\\
\\ 
\textbf{Class II}$^\S$& 
$ \left(
\begin{array}{cccccc}
 0 & 0 & 0 & \Lambda _{14} & 0 & \Lambda _{16} \\
 \Lambda _{21} & \Lambda _{22} & \Lambda _{23} & 0 & \Lambda _{25} & 0 \\
 0 & 0 & 0 & \Lambda _{34} & 0 & \Lambda _{36} \\
\end{array}
\right)$ &
\begin{tabular}{ccc}
\hmn{2}&\hmn{m}& \hmn{*2} 
\end{tabular} 
& 8 & D(2)+Q(3)+O(3)  \\
\\
\textbf{Class III}$^\S$ & $\left(
\begin{array}{cccccc}
 \Lambda _{11} & \Lambda _{12} & \Lambda _{13} & 0 & \Lambda _{15} & 0 \\
 0 & 0 & 0 & \Lambda _{24} & 0 & \Lambda _{26} \\
 \Lambda _{31} & \Lambda _{32} & \Lambda _{33} & 0 & \Lambda _{35} & 0 \\
\end{array}
\right)$ & 
\begin{tabular}{ccc}
\hmn{2'}& \hmn{m'}& \hmn{`{2'}} 
\end{tabular} 
& 10 & D(4)+Q(2)+O(4)  \\
\\
\textbf{Class IV}$^\S$ & $\left(
\begin{array}{cccccc}
 0 & 0 & 0 & 0 & \Lambda _{15} & 0 \\
 0 & 0 & 0 & \Lambda _{24} & 0 & 0 \\
 \Lambda _{31} & \Lambda _{32} & \Lambda _{33} & 0 & 0 & 0 \\
\end{array}
\right)$ &
\renewcommand{\arraystretch}{1.5}
\begin{tabular}{cc}
\hmn{2'2'2}&\hmn{m'm2'}\\
\hmn{m'm'2}&\hmn{m'm'm}
\end{tabular} & 5 & D(2)+Q+O(2) \\
\\
Class V$^\dagger$ & $\left(
\begin{array}{cccccc}
 0 & 0 & 0 & \Lambda _{14} & 0 & 0 \\
 0 & 0 & 0 & 0 & \Lambda _{25} & 0 \\
 0 & 0 & 0 & 0 & 0 & \Lambda _{36} \\
\end{array}
\right)$ &
\begin{tabular}{ccc}
\hmn{222}& \hmn{mmm}& \hmn{mm2}
\end{tabular} 
 & 3 & Q(2)+O  \\
\\
\textbf{Class VI}$^\S$ & $\left(
\begin{array}{cccccc}
 \Lambda _{11} & -\Lambda _{11} & 0 & \Lambda _{14} & \Lambda _{15} & -2 \Lambda _{22} \\
 -\Lambda _{22} & \Lambda _{22} & 0 & \Lambda _{15} & -\Lambda _{14} & -2 \Lambda _{11} \\
 \Lambda _{31} & \Lambda _{31} & \Lambda _{33} & 0 & 0 & 0 \\
\end{array}
\right)$ &
\begin{tabular}{cc}
\hmn{3}&\hmn{-3}
\end{tabular} 
& 6 & D(2)+Q+O(3)  \\
\\
\textbf{Class VII}$^\S$ & $\left(
\begin{array}{cccccc}
 0 & 0 & 0 & 0 & \Lambda _{15} & -2 \Lambda _{22} \\
 -\Lambda _{22} & \Lambda _{22} & 0 & \Lambda _{15} & 0 & 0 \\
 \Lambda _{31} & \Lambda _{31} & \Lambda _{33} & 0 & 0 & 0 \\
\end{array}
\right)$ &
\begin{tabular}{ccc}
\hmn{32'}&\hmn{3m'}&\hmn{-3m'}
\end{tabular} 
& 4 & D(2)+O(2)  \\
\\
\textit{Class VIII}$^*$& $\left(
\begin{array}{cccccc}
 \Lambda _{11} & -\Lambda _{11} & 0 & \Lambda _{14} & 0 & 0 \\
 0 & 0 & 0 & 0 & -\Lambda _{14} & -2 \Lambda _{11} \\
 0 & 0 & 0 & 0 & 0 & 0 \\
\end{array}
\right)$ &
\begin{tabular}{ccc}
\hmn{32}&\hmn{3m}&\hmn{-3m}
\end{tabular}
& 2 & Q+O  \\
\\
Class IX$^\dagger$ & $\left(
\begin{array}{cccccc}
 0 & 0 & 0 & \Lambda _{14} & \Lambda _{15} & 0 \\
 0 & 0 & 0 & -\Lambda _{15} & \Lambda _{14} & 0 \\
 \Lambda _{31} & -\Lambda _{31} & 0 & 0 & 0 & \Lambda _{36} \\
\end{array}
\right)$ &
\begin{tabular}{ccc}
\hmn{4'}&\hmn{-4'}&\hmn{*{4'}}
\end{tabular}
& 4 & Q(2)+O(2)  \\
\\
Class X$^\dagger$ & $\left(
\begin{array}{cccccc}
 0 & 0 & 0 & \Lambda _{14} & 0 & 0 \\
 0 & 0 & 0 & 0 & \Lambda _{14} & 0 \\
 0 & 0 & 0 & 0 & 0 & \Lambda _{36} \\
\end{array}
\right)$ & 
\begin{tabular}{cc}
\hmn{4'22'}&\hmn{-4'2m'}
\end{tabular}
& 2 & Q+O \\

Class XI$^\dagger$ & $\left(
\begin{array}{cccccc}
 0 & 0 & 0 & 0 & \Lambda _{15} & 0 \\
 0 & 0 & 0 & -\Lambda _{15} & 0 & 0 \\
 \Lambda _{31} & -\Lambda _{31} & 0 & 0 & 0 & 0 \\
\end{array}
\right)$ &
\begin{tabular}{ccc}
\hmn{-4'2'm}&\hmn{4'm'm}&\hmn{*{4'}m'm}
\end{tabular}
& 2 & Q+O  \\
\\
\textit{Class XII}$^*$ & $\left(
\begin{array}{cccccc}
 \Lambda _{11} & -\Lambda _{11} & 0 & 0 & 0 & -2 \Lambda _{22} \\
 -\Lambda _{22} & \Lambda _{22} & 0 & 0 & 0 & -2 \Lambda _{11} \\
 0 & 0 & 0 & 0 & 0 & 0 \\
\end{array}
\right)$ &
\begin{tabular}{ccc}
\hmn{6'}&\hmn{-6'} & \hmn{`{6'}}
\end{tabular}
 & 2 & O(2)  \\
\\
\textbf{Class XIII}$^\S$ & $\left(
\begin{array}{cccccc}
 0 & 0 & 0 & \Lambda _{14} & \Lambda _{15} & 0 \\
 0 & 0 & 0 & \Lambda _{15} & -\Lambda _{14} & 0 \\
 \Lambda _{31} & \Lambda _{31} & \Lambda _{33} & 0 & 0 & 0 \\
\end{array}
\right)$ &
\renewcommand{\arraystretch}{1.5}
\begin{tabular}{cccc}
(a)&\hmn{6}&\hmn{-6}&\hmn{*6}\\
 (b)&\hmn{4}&\hmn{-4}&\hmn{*4} 
\end{tabular}
& 4 & D(2)+Q+O  \\
\\
\textit{Class XIV}$^*$ & $\left(
\begin{array}{cccccc}
 \Lambda _{11} & -\Lambda _{11} & 0 & 0 & 0 & 0 \\
 0 & 0 & 0 & 0 & 0 & -2 \Lambda _{11} \\
 0 & 0 & 0 & 0 & 0 & 0 \\
\end{array}
\right)$ & 
\renewcommand{\arraystretch}{1.5}
\begin{tabular}{ccc}
\hmn{6'22'}&\hmn{-6'2m'}&\hmn{-6'm2'}\\
\hmn{6'mm'}&\hmn{`{6'}mm'}
\end{tabular} 
 & 1 & O  \\
\\
\textit{Class XV}$^*$ & $\left(
\begin{array}{cccccc}
 0 & 0 & 0 & \Lambda _{14} & 0 & 0 \\
 0 & 0 & 0 & 0 & -\Lambda _{14} & 0 \\
 0 & 0 & 0 & 0 & 0 & 0 \\
\end{array}
\right)$ & 
\renewcommand{\arraystretch}{1.5}
\begin{tabular}{ccccc}
(a)&\hmn{622}& \hmn{-6m2} & \hmn{6mm}& \hmn{*6mm}\\
(b)& \hmn{422}&\hmn{-42m}&\hmn{4mm}&\hmn{*4mm}
\end{tabular} 
& 1 & Q  \\
\\
\textbf{Class XVI}$^\S$ & $\left(
\begin{array}{cccccc}
 0 & 0 & 0 & 0 & \Lambda _{15} & 0 \\
 0 & 0 & 0 & \Lambda _{15} & 0 & 0 \\
 \Lambda _{31} & \Lambda _{31} & \Lambda_{33} & 0 & 0 & 0 \\
\end{array}
\right)$ & 
\renewcommand{\arraystretch}{1.5}
\begin{tabular}{ccccc}
(a)&\hmn{62'2'}&\hmn{-6m'2'}&\hmn{6m'm'}&\hmn{*6m'm'}\\
(b)&\hmn{42'2'}&\hmn{-42'm'}&\hmn{4m'm'}&\hmn{*4m'm'}
\end{tabular} 
& 3 & D(2)+O  \\
\\
\\
\textit{Class XVII}$^*$ & $\left(
\begin{array}{cccccc}
 0 & 0 & 0 & \Lambda _{14} & 0 & 0 \\
 0 & 0 & 0 & 0 & \Lambda _{14} & 0 \\
 0 & 0 & 0 & 0 & 0 & \Lambda _{14} \\
\end{array}
\right)$ &
\renewcommand{\arraystretch}{1.5}
\begin{tabular}{cccc}
(a)& \hmn{4'32'}&\hmn{-4'3m'}&\hmn{m-3m'} \\
(b)&\hmn{23}&\hmn{m-3}
\end{tabular}
 & 1 & O  \\
 \\
 \hline
 \\
 \textit{Class XVIII}$^*$ & $ 
 B^{eff}_i=\Lambda (k) \epsilon_{ijl}(k_j^3k_l-k_l^3k_j)
 $ &
\renewcommand{\arraystretch}{1.5}
\begin{tabular}{ccc}
\hmn{432}&\hmn{-43m}&\hmn{m-3m}
\end{tabular}
 & 1 & H  \\
\\

\hline\hline\hline

\\

\end{xtabular*}
\begin{tabular}{l}
(\S) Classes allowing a magnetic dipole term (Type-I altermagnets in ref. \onlinecite{Cheong2024})\\
($\dagger$) Classes allowing spin splitting along a permitted collinear AFM direction (`strong-collinear' altermagnetic classes)\\
(*) Classes not allowing spin splitting along a permitted collinear AFM direction (`weak-collinear' altermagnetic classes)\\
\end{tabular}
\end{tiny}
\end{widetext} 
}

\section{Connection with spin groups: the cubic symmetry}
\label{sec: spin_group_cubic}

In several important cases (both collinear and non-collinear), the crystal symmetry is only broken by the direction of the magnetic moments.  A classic case is that of hematite (Fe$_2$O$_3$),\cite{Morin} where the 3-fold symmetry is broken at room temperature but remains unbroken below the so-called Morin transition ($\approx$260K).  Since the magneto-elastic interaction is often small, one needs to take into account the effect of the approximate crystal symmetry (termed `pseudo-symmetry' in the remainder) on the spin textures.  In the case of collinear structures, this can be done effectively using SGs --- a treatment that has been presented in previous literature. \cite{Smejkal2022, Smejkal2022b} One needs to establish how the approach based on MPG needs to be modified to take pseudo-symmetry into account, so that the two approaches (MPG and SG) are fully connected and consistent.

Although he effective field $\vec{B}^{eff}$ along the staggered magnetisation can always be identified from the MPG treatment,  pseudo-symmetries introduce additional (approximate) constraints that are not immediately apparent from the MPG (see sec. \ref{sec: ex_spingroups} for a complete example).  A simple and effective technique to address this issue is to is to rotate all the magnetic moments into a high-symmetry direction, so that as many pseudo-symmetry operators as possible become re-established as exact symmetries in the MPG.  When the magneto-elastic interaction is weak, such high-symmetry phase is almost always physical and, if it is not the ground state, it can usually be reached by applying a relatively small magnetic field.  However, this approach is not always viable, particularly in cubic symmetry, because, for collinear phases, cubic crystal symmetry can never be fully restored by any direction of the staggered magnetisation. These are precisely the cases in which a separate symmetry analysis using SGs is most necessary. 


Here, SG analysis is implemented within the same tensorial framework as the MPGs and by using using Shubnikov groups rather than the (isomorphic) binary SG.  For collinear structures, this is equivalent to treating textures in both real and momentum space as \emph{time-reversal-odd scalars} (see Appendix \ref{App: spin_groups}).  Treating real-space collinear magnetic textures in this manner is by no means unprecedented --- in fact, classifying magnetic structures based on scalar ordering patterns (e.g., $A$-type, $C$-type, $G$-type, etc.) goes back all the way to the very beginning of the experimental research on AFMs (ref. \onlinecite{Wollan1955} fig. 18).  Scalar patterns also play a key role in the theory of exchange multiplets, \cite{Izyumov1980} which, crucially, can also be applied to non-collinear structures.  Within this framework, a given collinear magnetic structure will then come to be associated with two distinct Shubnikov groups: the Shubnikov MPG, which describes the exact symmetry of the combined crystal and axial-vector magnetic structures, and the Shubnikov SG, which describes the symmetry of the corresponding scalar magnetic ordering pattern.  In the absence of pseudo-symmetries, the two Shubnikov groups will act in the same way on the crystal structure (i.e., they have the same associated `grey' group) and can be readily obtained from one another if the direction of the magnetic moment is known.  However, in the presence of pseudo-symmetries, the Shubnikov SG will generally have a higher symmetry,  the additional operators corresponding precisely to the pseudo-symmetries.   By construction, Shubnikov (or binary) SG can be employed to classify and delimit any magnetic property that does not depend on the direction of the magnetic moment, and, in the present context, can be used to construct symmetrised scalar-field tensors (polynomial forms) in momentum space at any order.


Lowest-order polynomial forms for all cubic SGs have been obtained using scalar-field tensors, as discussed in sec. \ref{sec: Jahn}, and are listed in Table \ref{Tab: cubic_forms} together with the corresponding MPGs for magnetic moments along the three cubic symmetry directions $[001]$, $[110]$ and $[111]$.  When comparing these forms with the corresponding strong-collinear forms of the same MPGs, one notices that the former can be obtained from the latter by fixing some of the parameters to have specific values.  In other words, the additional approximate symmetry implied by the scalar textures (pseudo-symmetry) manifests itself by establishing a link between effective-field parameters that would otherwise be independent.  This becomes clear when considering specific cubic symmetries.

The cubic SGs $\hmn{23}$, $\hmn{m-3}$, $\hmn{432}$ $\hmn{-43m}$ and $\hmn{m-3m}$ are all ferro- (or ferri-) magnetic, and the lowest-order cubic polynomial form is completely isotropic.  The corresponding MPGs all admit a dipole component, which is, however, not necessarily isotropic.  For example, the polynomial form for MPG $\hmn{42'2'}$ is $\Lambda_1(k) (k_x^2+k_y^2)+\Lambda_2(k) k_z^2$.  Therefore, the pseudo-cubic symmetry had the result of imposing $\Lambda_1(k) =\Lambda_2(k)\; \forall k$, as stated.

The case of the three remaining cubic groups $\hmn{\tilde{4}3\tilde{2}}$, $\hmn{\tilde{{-4}}3\tilde{m}}$ and $\hmn{m -3 \tilde{m}}$ is considerably more interesting, and is discussed in more detail in sec. \ref{sec: cubic_groups}.  The lowest-order cubic polynomial form is of rank 7 (rank 6 for the corresponding SGs), while the corresponding MPGs admit lower-order polynomial forms (see tab. \ref{Table_1} and \ref{Tab: newTab}).  As always, these lower-order MPG forms are contained in the rank-7 forms --- for example, the trigonally-symmetric (Class VIII) rank-5 form $\Lambda(k) (k_x^2 - 3 k_y^2)k_x k_z$ has corresponding rank-7 components $ (k_x^2 - 3 k_y^2)k_x k_z \left(\Lambda_1(k)(k_x^2+k_y^2)+\Lambda_2(k)k_z^2\right)$, since $(k_x^2+k_y^2)$ and $k_z^2$ are totally symmetric in trigonal symmetry.  However, in the presence of cubic pseudo-symmetry, these components are linked together and with other rank-7 components, yielding the pseudo-cubic rank-7 form displayed in tab. \ref{Tab: cubic_forms}. All other trigonal components only arise from the symmetry breaking due to the magnetic moment direction, and are expected to be small for weak magneto-elastic interactions/SOC.

\begin{table*}[h!]
\centering
\caption{\label{Tab: newTab} Polynomial forms of the effective field component $z$, which in the listed classes is along the collinear staggered magnetisation (strong-collinear altermagnetism).  Classes listed here are weak-collinear altermagnetic at the rank-3 (quadratic) level (see tab. \ref{Table_1}).}
\begin{tabular}{|c|c|c|}
\hline
Class &Rank & $B_z^{eff} (\vec{k})$\\
\hline
VIII & 5&$\Lambda(k)  (k_x^2 - 3 k_y^2 ) k_x k_z $\\

XII & 5&$ \left(\Lambda_1(k)  (3 k_x^2 k_y - k_y^3 ) - \Lambda_2(k) (3 k_x k_y^2  - k_x^3 ) \right)k_z$\\
XIV &5 &$ \Lambda(k) ( k_x^3-3 k_y^2 k_x ) k_z$\\

XVa &7&$ \Lambda (k)  (3 k_x^4 - 10 k_x^2 k_y^2 + 3 k_y^4)k_x k_y$\\

XVb &5&$\Lambda (k) (k_x^3k_y - k_y^3k_x)$\\
\hline
\hline
\end{tabular}
\end{table*}

\section{Examples}

\label{sec: examples}

Here, I provide a set of examples to illustrate the practical use of this method and in particular of Table \ref{Table_1}, and to highlight some of the issues arising from different crystallographic conventions.  All the examples and references were generated using the program MAGNDATA from the Bilbao Crystallographic Server. \cite{Gallego2016}  The figures display the `texture pattern' as a normalised $\vec{B}^{eff}$ vector field, where the normalisation function is $1/(k_x^2+k_y^2+k_z^2)^{(n-1)/2}$ where $n$ is the rank of the tensor.  These `texture patterns'  correspond to the gnomonic projection of the textures, i.e., a projection from the centre of the unit sphere to a plane tangent to it.  This projection is most convenient to display the symmetry (it preserves angles at the centre) and the details of the textures.  In these figures, $k_x$ and $k_y$ are in arbitrary dimensionless units, and the conversion to spherical coordinates is $\tan \theta = \sqrt{k_x^2+k_y^2}$, $\tan \phi=k_y/k_x$.  There is a close correspondence between these texture patterns and those displayed schematically in fig. 2 of \v{S}mejkal \textit{et al.}\cite{Smejkal2022} (more details in the figure legends). An example of texture plotted on the surface of a sphere in momentum space is shown in sec. \ref{sec: cubic_groups}.

\subsection{Class XIV (weak-collinear at rank 3)}
Class XIV comprises five hexagonal MPGs: 
$\hmn{`{6'}mm'}$, $\hmn{-6'2m'}$ (a MPG adopted, for example, by TmAgGe, ref. \onlinecite{Baran2009}), $\hmn{-6'm2'}$, $\hmn{6'mm'}$ (HoMnO$_3$, MSG $\hmn{P6'_3cm'}$, ref. \onlinecite{Brown2006}) and $\hmn{6'22'}$.
  The expression for  $\vec{B}^{eff}(\vec{k})$ for this class at the rank-3 tensor level is:

\begin{eqnarray}
B_x^{eff} (\vec{k})&=&\Lambda_{11}(k)(k_x^2-k_y^2)  \nonumber\\
B_y^{eff} (\vec{k})&=&\Lambda_{11}(k) ( -2 k_x k_y )\nonumber\\
B_z^{eff} (\vec{k})&=&0
\end{eqnarray}

the spin texture being parametrised by a \emph{single} scalar function of $k$, $\Lambda_{11}(k)$.  The spin quantisation axis is in the $xy$ plane and therefore always perpendicular to the high-symmetry direction, i.e., the allowed direction for the N\'eel vector.  The spin texture is parallel to $\vec{k}$ along the six directions $\Gamma - K$ and $\Gamma - K'$,  while it is perpendicular to $\vec{k}$ along the $\Gamma - M$ lines (see ref. \onlinecite{Bradley2010} for the BZ point notation).  Since the three $K$ points and the three $K'$ points are equivalent, $\Lambda_{14}(k)$ must be $=0$ both at the $\Gamma$ point and at $k=|K|$.  There is no additional constraint at points $M$.


Class XIV is also useful to illustrate how to deal with different axes orientations, which is essential to employ Table \ref{Table_1} correctly.  As a preamble, one should remark that, in the presence of a crystal lattice, the orientation of the point group directions is not arbitrary, but is related to that of the crystal axes.  So, for example, symbols $\hmn{-62m}$ and $\hmn{-6m2}$ refer to the same point group, being related by a 90$^\circ$ rotation of the in-plane axes.  However, $\hmn{P-62m}$ and $\hmn{P-6m2}$ are distinct space groups, since the in-plane directions are now linked to the crystal axes.

 The symbols, $\hmn{6'2'2}$, $\hmn{-6'm'2}$, $\hmn{-6'2'm}$ (e.g., ThMn$_2$,  ref. \onlinecite{Deportes1987} and CsFeCl$_3$, ref. \onlinecite{Hayashida2018}, both with MSG, $\hmn{P-6'2'm}$,) $\hmn{6'm'm}$ (e.g., YbMnO$_3$, MSG $\hmn{P6'_3c'm}$, ref. \onlinecite{Fabreges2008} ), and $\hmn{`{6'}m'm}$ (e.g., CrSb, MSG $\hmn{P`{6'_3}m'c}$, ref. \onlinecite{Yuan2020b}) refer to the \emph{same} point groups ($\hmn{6'22'}$, $\hmn{-6'2m'}$, $\hmn{-6'm2'}$, $\hmn{6'mm'}$, and $\hmn{`{6'}mm'}$, respectively), with the in-plane axes rotated by 90$^\circ$. Therefore, when dealing with MSGs such as $\hmn{P-6'2'm}$ or $\hmn{P`{6'_3}m'c}$, one should employ the \emph{rotated form of the tensor}: \footnote{I did not create a separate class for these symbols, since this tensor is obviously related to that in Table \ref{Table_1} and in order to avoid proliferation of classes. However, the reader should be advised that simple transformations such as the one in  eq. \ref{eq: tens_trans_vers} may be necessary, not only in this case but also to deal with non-standard conventions (e.g., $2'2'2$ vs  $22'2'$, etc. in Class IV).}

 \begin{equation}
 \label{eq: tens_trans_vers}
 \left(
\begin{array}{cccccc}
0&0& 0 & 0 & 0 & -2 \Lambda _{22} \\
- \Lambda _{22} & \Lambda _{22}  & 0 & 0 & 0 & 0 \\
 0 & 0 & 0 & 0 & 0 & 0 \\
\end{array}
\right) 
 \end{equation}

The expression for  $\vec{B}^{eff}(\vec{k})$ for this set being:

\begin{eqnarray}
B_x^{eff} (\vec{k})&=&\Lambda_{11}(k) ( -2 k_x k_y )\nonumber\\
B_y^{eff} (\vec{k})&=&\Lambda_{11}(k)(k_y^2-k_x^2)  \nonumber\\
B_z^{eff} (\vec{k})&=&0
\end{eqnarray}

Among the compounds listed above, TmAgGe, HoMnO$_3$,  YbMnO$_3$, CsFeCl$_3$ and ThMn$_2$ are reported to have non-collinear magnetic structures, while SbCr is reportedly collinear.

As discussed in sec. \ref{sec: higher_orders}, Class XIV becomes strong-collinear at the rank-5 (quartic) level, the effective field along the $z$ axis (the allowed direction for the collinear staggered magnetisation) being (see tab. \ref{Tab: newTab}):

\begin{equation}
B^{eff}_z(\vec{k})=\Lambda(k) ( k_x^3-3 k_y^2 k_x ) k_z
\end{equation}

\begin{figure}[!h]
\centering
\includegraphics[scale=0.5, trim={0 3cm , 0, 3cm},clip]{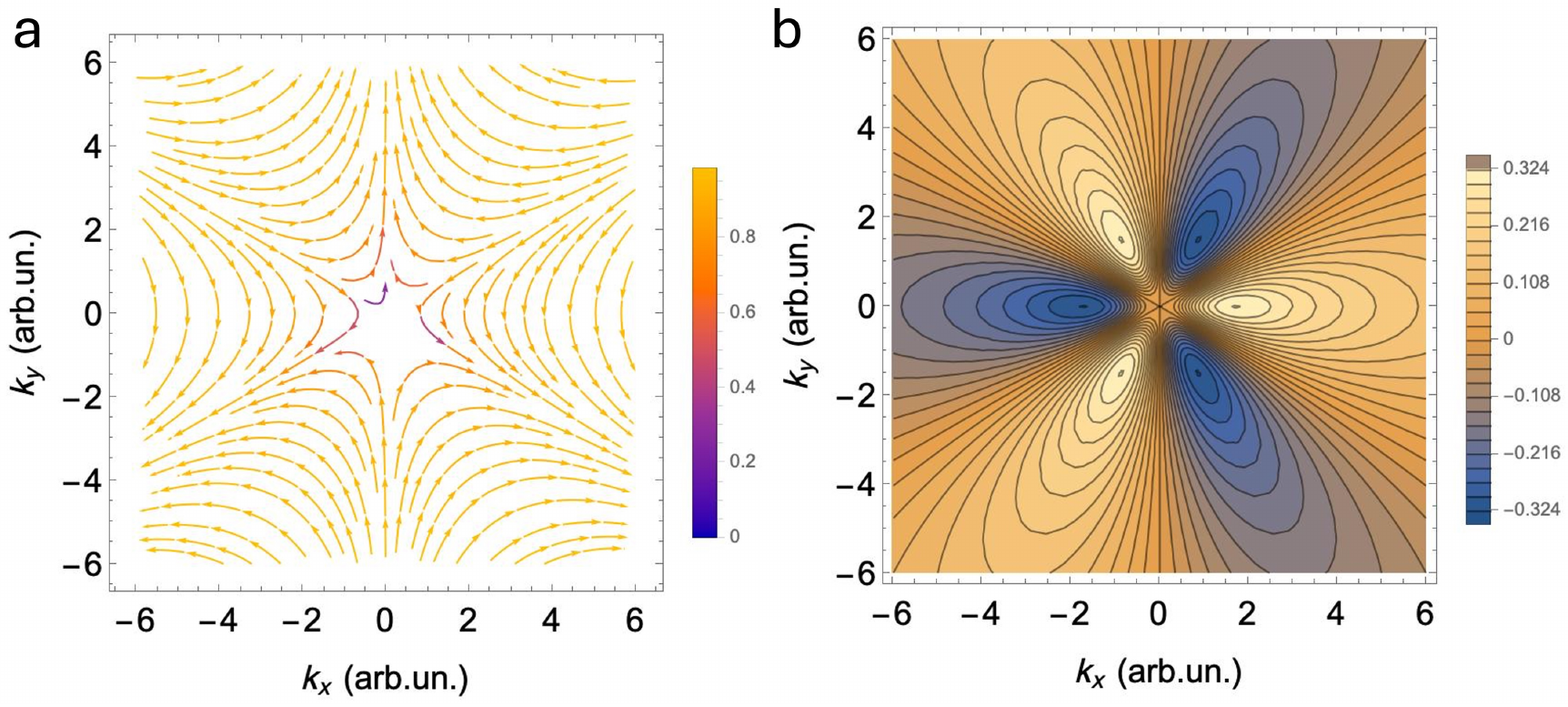}
\caption{ \label{fig: 6'22'}  In-plane, rank-3 (a) and out-of-plane, rank 5 (b) components of the `texture pattern' (normalised $\vec{B}^{eff}$ vector field, corresponding to the gnomonic projection of the spin texture --- see text), for Class XIV (e.g., MPG $\hmn{6'22'}$). The normalisation functions are $1/(k_x^2+k_y^2+k_z^2)^{(n-1)/2}$, $n$ being the rank of the tensor.  Contours were plotted for $k_z=1$ and are symmetric/antisymmetric, respectively for panels (a)/(b), by exchange $k_z\rightarrow -k_z$. Panel (b) to be compared with fig. 2, sub-panel B4 of ref. \onlinecite{Smejkal2022}.}
\end{figure}

Fig. \ref{fig: 6'22'} displays the corresponding texture patterns.  

\subsection{Classes X and XI (strong-collinear at rank 3)}

Classes X and XI are closely related, \cite{Note4} because their tensor forms are related by a $45^\circ$ rotation.  However, these forms look sufficiently different to make distinct classes useful in practice.  The difference between the two classes stems from the fact that, in Class X, operators along the the primary ($[001]$) and tertiary ($[110]$) symmetry directions are primed while, for Class XI, operators along the the primary ($[001]$) and secondary ($[100]$) symmetry directions are primed.

The tensor form for Class X  allows spin splitting along all three axes.  However, setting $\Lambda_{14}\approx0$, one finds that the spin quantisation axis can be predominantly parallel or antiparallel to the $z$ axis, i.e., the allowed N\'eel vector direction for collinear antiferromagnetism.  The maximum effective Zeeman field along the $z$ axis is for wavevectors in the $(110)$ and $(1\bar{1}0)$ directions, and the sign of the field switches between these two directions.  

Class X comprises the MPGs $\hmn{-4'2m'}$ (e.g., Pb$_2$MnO$_4$, ref. \onlinecite{Kimber2007}) and $\hmn{4'22'}$ (Er$_2$Ge$_2$O$_7$, ref. \onlinecite{Taddei2019}), but also those corresponding to the non-standard setting $\hmn{-4'2'm}$, $\hmn{4'mm'}$ and $\hmn{*4'mm'}$ (e.g., LiFe$_2$F$_6$, ref. \onlinecite{Shachar1972}, and RuO$_2$, ref. \onlinecite{Berlijn2017}, both with MSG $\hmn{P*{4'_2}nm'}$) .   The characteristic spherical tensor expansion for this class is:

\begin{equation}
A(k) \,\ten{Q}^{III}+B(k)\, \ten{O}^{I}
\end{equation}

where $\ten{Q}^{III} $ and $\ten{O}^{I} $ are fixed tensors defined in Table \ref{Table_2}.  The spin texture is:

\begin{eqnarray}
\label{eq: tetr_field_X}
B_x^{eff} (\vec{k})&=&\Lambda_{14}(k) k_y k_z  \nonumber\\
B_y^{eff} (\vec{k})&=& \Lambda_{14}(k) k_x k_z \nonumber\\
B_z^{eff} (\vec{k})&=&\Lambda_{36}(k) k_x k_y
\end{eqnarray}

\begin{figure}[!h]
\centering
\includegraphics[scale=0.5, trim={0 3cm , 0, 3cm},clip]{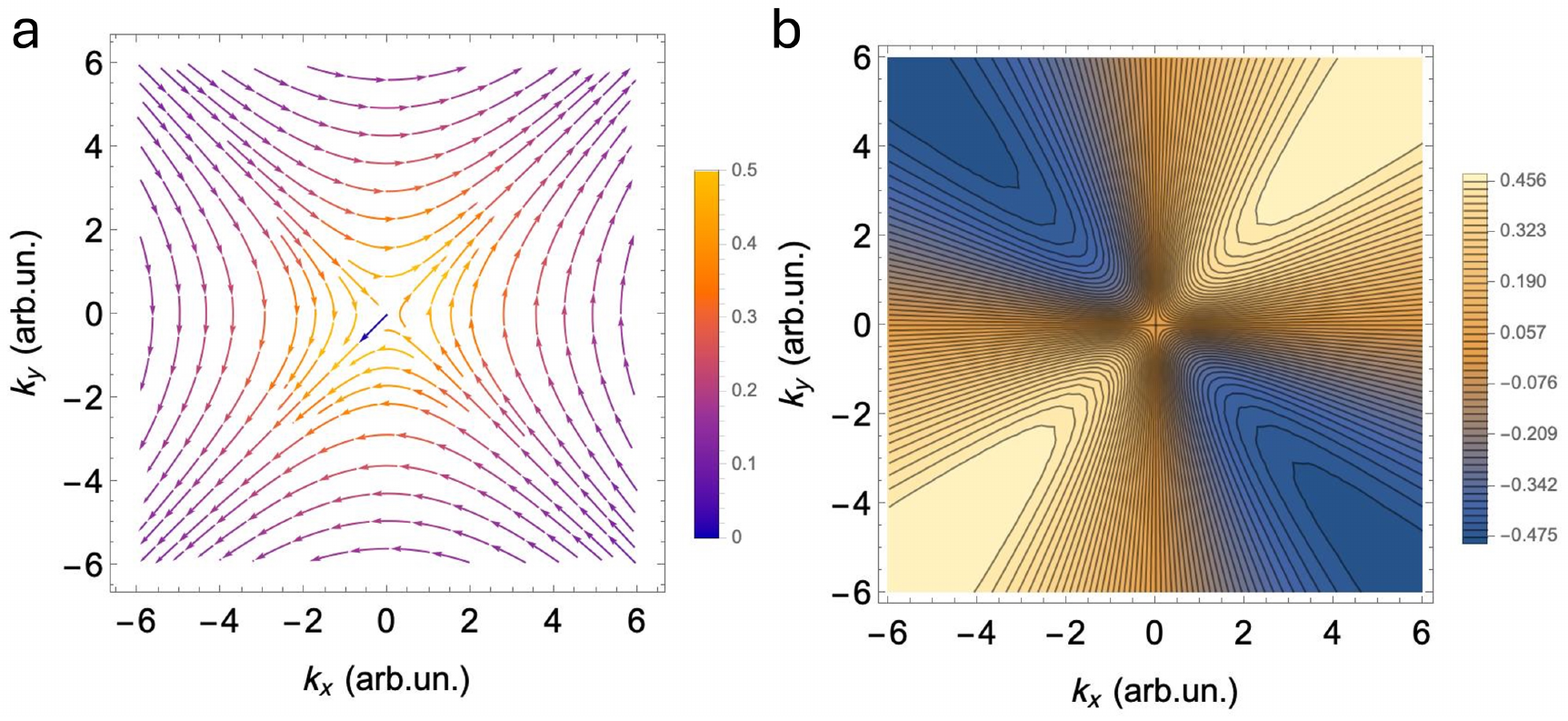}
\caption{ \label{fig: 4'22'}  In-plane (a) and out-of-plane (b) components (both of rank 3) of the texture pattern for Class X (e.g., MPG $\hmn{4'22'}$). The normalisation function is $1/(k_x^2+k_y^2+k_z^2)$.  Contours were plotted for $k_z=1$ and are antisymmetric/symmetric, respectively for panels (a)/(b), by exchange $k_z\rightarrow -k_z$. }
\end{figure}

In the primitive tetragonal cell (adopted by Pb$_2$MnO$_4$, Er$_2$Ge$_2$O$_7$, LiFe$_2$F$_6$ and RuO$_2$), the four $M$ points have reciprocal-space coordinates $(\frac{1}{2}, \frac{1}{2}, 0)$, $(-\frac{1}{2}, \frac{1}{2}, 0)$, $(\frac{1}{2}, -\frac{1}{2}, 0)$ and $(-\frac{1}{2}, -\frac{1}{2}, 0)$ and are all related by reciprocal lattice vectors.  However, according to eq. \ref{eq: tetr_field_X}, $(\frac{1}{2}, \frac{1}{2}, 0)/(\frac{1}{2}, -\frac{1}{2}, 0)$ and $(-\frac{1}{2}, \frac{1}{2}, 0)/(\frac{1}{2}, -\frac{1}{2}, 0)$ must have opposite spin textures, which means that $\Lambda_{36}(k)=0$ for $k=k_M$.  The in-plane and out-of-plane spin-texture patterns, both of rank 3, are displayed in fig. \ref{fig: 4'22'} .

Class XI comprises the standard-setting MPGs $\hmn{-4'2'm}$, $\hmn{4'm'm}$, $\hmn{*{4'}m'm}$ (e.g., the pyrochlores Er$_2$Ti$_2$O$_7$ and Er$_2$Ru$_2$O$_7$, with MSG $\hmn{I~{4'_1}m'd}$, refs. \cite{Taira2003,Poole2007}), and also the non-standard-settings $\hmn{-4'2'm}$ and $\hmn{4'2'2}$.  The characteristic spherical tensor expansion for this class is:

\begin{equation}
A(k) \,\ten{Q}_z^{I}+B(k)\, \ten{O}_z^{II}
\end{equation}

where $\ten{Q}_z^{I} $ and $\ten{O}_z^{II} $ are fixed tensors defined in Table \ref{Table_2}.  The spin texture is rotated by 45$^\circ$ with respect to Class X:

\begin{eqnarray}
\label{eq: tetr_field_XI}
B_x^{eff} (\vec{k})&=&\Lambda_{15}(k) k_x k_z  \nonumber\\
B_y^{eff} (\vec{k})&=&- \Lambda_{15}(k) k_y k_z \nonumber\\
B_z^{eff} (\vec{k})&=&\Lambda_{13}(k) (k_x^2- k_y^2)
\end{eqnarray}


Among the compounds listed above, Pb$_2$MnO$_4$ Er$_2$Ge$_2$O$_7$, Er$_2$Ti$_2$O$_7$ are reported to have non-collinear magnetic structures, while LiFe$_2$F$_6$, Er$_2$Ru$_2$O$_7$, and RuO$_2$ are reportedly collinear (but doubts have recently be raised concerning magnetism in RuO$_2$ --- see ref. \onlinecite{Kessler2024}).

\subsection{Class XV: higher-order strong-collinear altermagnetism}
\label{sec: Class XV}

Class XV is weak-collinear at the rank-3 level, with 

\begin{eqnarray}
\label{eq: tetr_field_XV}
B_x^{eff} (\vec{k})&=&\Lambda_{14}(k) k_y k_z  \nonumber\\
B_y^{eff} (\vec{k})&=&- \Lambda_{14}(k) k_x k_z \nonumber\\
B_z^{eff} (\vec{k})&=&0
\end{eqnarray}

\begin{figure}[!h]
\centering

\includegraphics[scale=0.5, trim={0 3cm , 0, 3cm},clip]{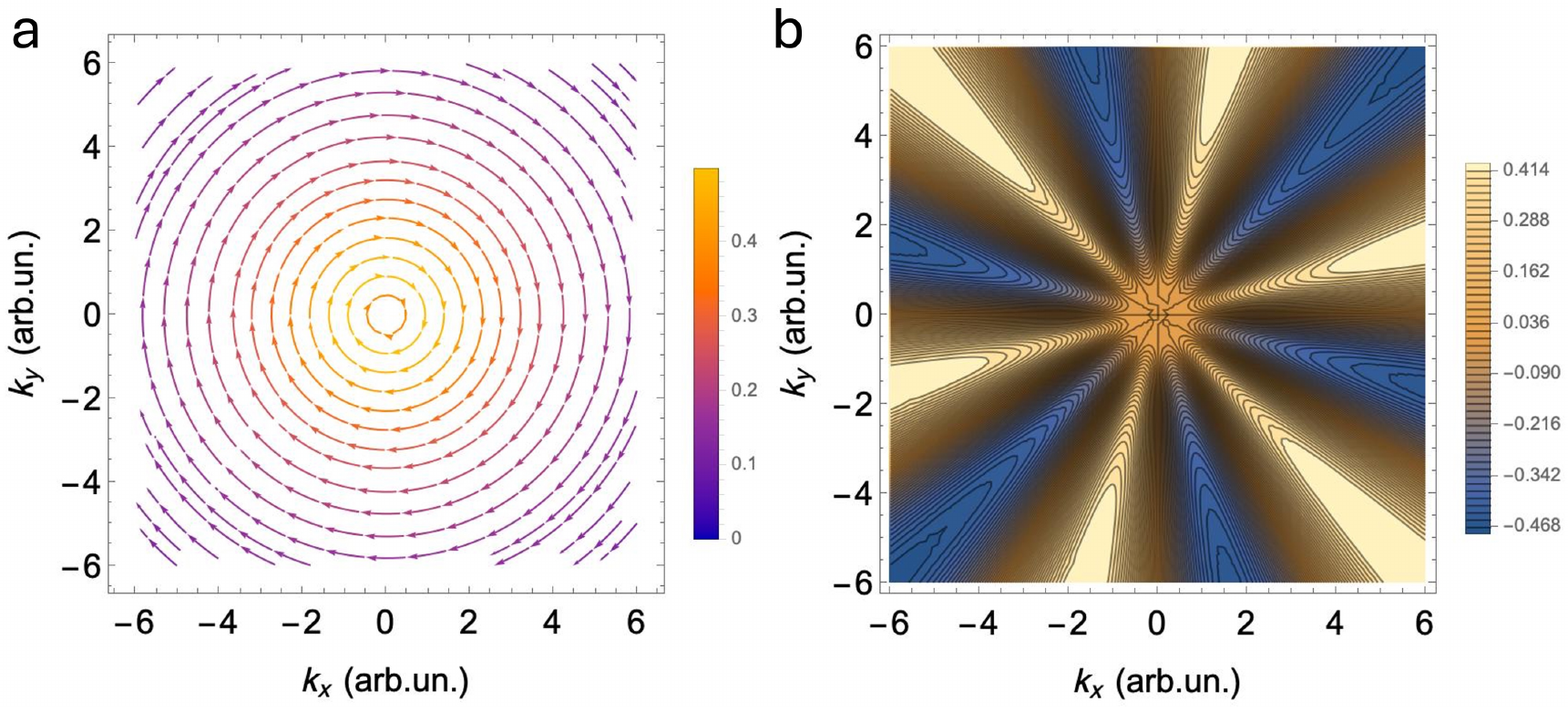}
\caption{ \label{fig: 622}  In-plane, rank-3 (a) and out-of-plane, rank-7 (b) components of the texture pattern for Class XVa (e.g., MPG $\hmn{622}$). The normalisation function is as in previous figures.  Contours were plotted for $k_z=1$ and are antisymmetric/symmetric, respectively for panels (a)/(b), by exchange $k_z\rightarrow -k_z$. Panel (b) to be compared with fig. 2, sub-panel P6 of ref. \onlinecite{Smejkal2022}.}
\end{figure}

\begin{figure}[!h]
\centering
\includegraphics[scale=0.5, trim={0 3cm , 0, 3cm},clip]{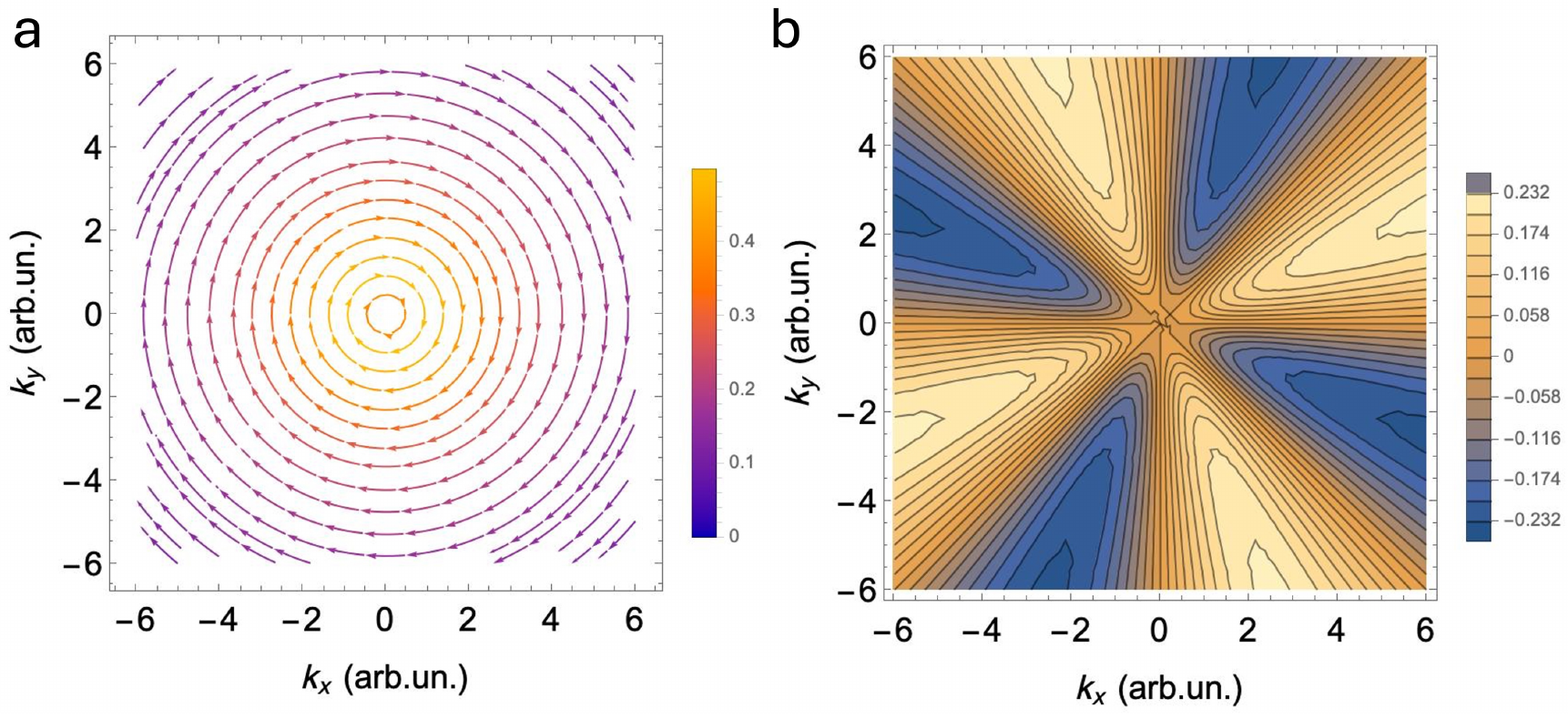}
\caption{ \label{fig: 422}  In-plane, rank-3 (a) and out-of-plane, rank-5 (b) components of the texture pattern  for Class XVb (e.g., MPG $\hmn{422}$). The normalisation function is as in previous figures.  Contours were plotted for $k_z=1$ and are antisymmetric/symmetric, respectively for panels (a)/(b), by exchange $k_z\rightarrow -k_z$.  Panel (b) to be compared with fig. 2, sub-panel P4 of ref. \onlinecite{Smejkal2022}.}
\end{figure}

Subclasses XVa and XVb become strong-collinear at the rank-7 and rank-5  level, respectively, the corresponding $B_z^{eff}$ components being listed in tab. \ref{Tab: newTab}.  Figs. \ref{fig: 622} and \ref{fig: 422} display the corresponding texture patterns.


\subsection{Spin-flop transitions: connection with spin groups}
\label{sec: ex_spingroups}


As discussed in sec. \ref{sec: spin_group_cubic}, pseudo-symmetries are often created across a spin flop transition, when spins are rotated away from a high-symmetry direction.  These examples are very useful to illustrate the connection between the MPG and SG treatment of the same collinear structure.

Let us consider the magnetic structure of CoF$_2$ (space group $\hmn{P*{4_2}nm}$), a well known piezomagnetic material.  \cite{Dzialoshinskii1957, Borovik-Romanov1960}  In zero applied magnetic field, CoF$_2$ possesses a fully compensated collinear structure with spins along the $c$-axis. In this compound, Co$^{2+}$ (3d$^7$) is strongly anisotropic, and a spin flop transition was reported at a magnetic field of 7 T, \cite{Wu2016}  which is typical of many 3d transition metal compounds. \footnote{Half-filled-shell ions such as Fe$^{3+}$ have smaller anisotropies, but even in such cases spin-lattice coupling and the Dzyaloshinskii-Moriya interaction cannot be neglected.} The MSG in the zero-field phase is $\hmn{P*{4'_2}nm'}$, and the MPG is $\hmn{*{4'}mm'}$ (Class X).  In the high-field phase and assuming magnetic moments along the $[100]$ or $[110]$ directions, the MPG is $\hmn{m'mm'}$, where the first $m'$ is perpendicular to the N\'eel vector direction (here chosen to be the $x$ axis), while the second $m'$ is perpendicular to the $z$ axis.  This MPG is a member of Class IV, with $y$ and $z$ exchanged.  Using the conventions indicated here above, we obtain:

\begin{eqnarray}
\ten{T}_{S \parallel c}&=&\left(\begin{array}{cccccc}
 0 & 0 & 0 & \Lambda _{14} & 0 & 0 \\
 0 & 0 & 0 & 0 & \Lambda _{14} & 0 \\
 0 & 0 & 0 & 0 & 0 & \Lambda _{36} \\
\end{array}
\right)\nonumber\\
\ten{T}_{S \parallel a}&=&\left(
\begin{array}{cccccc}
 0 & 0 & 0 & 0 &  0 & \Lambda _{15} \\
  \Lambda _{31} & \Lambda _{33} & \Lambda _{32} & \Lambda _{24} & 0 & 0 \\
 0&0&0& 0 & 0 & 0 \\
\end{array}
\right)
\end{eqnarray}

where, for $\ten{T}_{S \parallel a}$, I have exchanged $y$ and $z$ with respect to the standard setting in Table \ref{Table_1}.  The zero-field magnetic structure $\ten{T}_{S \parallel c}$ is strong-collinear for the collinear AFM direction along the $z$ axis, via the tensor element $\Lambda _{36} $.  When $\Lambda _{14} $ is set to zero, $\vec{B}^{eff}$ is also parallel to the $z$ axis and has the form $\Lambda_{36} k_x k_y$, being maximum when $\vec{k}$ is in the $xy/x\bar{y}$ directions.  The high-field magnetic structure would allow weak FM, as indicated by the presence of dipole terms in Class IV.  However, $\ten{T}_{S \parallel a}$ is also strong-collinear in the $x$ direction via the $\Lambda _{15} $ element.  When all other elements are set to zero, $\vec{B}^{eff}$ is along the $x$ direction (the direction of the AFM moments) and has the form $\Lambda_{24} k_y k_z$, being maximum when $\vec{k}$ is in the $yz/y\bar{z}$ directions.  This is exactly what one would expect based on the SG approach (see below), i.e., spin splitting along the AFM spin direction and with a reciprocal space pattern that does not depend on the spin direction.  The MPG approach correctly describes this and also includes other tensor elements that are allowed by symmetry (in this case along the $y$ direction), which are likely to be small for collinear structures, but may well be large for non-collinear structures.

Regardless of the spin orientation, the spin space group is $\hmn{P*{\tilde{4}_2}\tilde{n}m}$, where the ``$\tilde{\;\;}$'' symbol indicates spin flip in the SG sense, i.e., with the space-group operators only acting on atoms and not on spins.  The (point) SG is $\hmn{*{\tilde{4}}\tilde{m}m}$, and the scalar texture tensor, generated using MTENSOR is:

\begin{equation}
\ten{T}_{SG}=\left(\begin{array}{ccc}
 0 & \Lambda_{12}  & 0 \\
 \Lambda_{12}  & 0 & 0  \\
 0 & 0 & 0\\
\end{array}\right)
\end{equation}

The corresponding texture has the form $\Lambda_{12} k_x k_y$, which is identical to the previous ones with an appropriate exchange of axes.

In the example just illustrated and for both directions of the N\'eel vector, the MPG and SG treatments yield identical momentum-space textures projected along the N\'eel vector. In addition, the MPG treatment evidences other `weak' components of the textures in different directions, including a `weak FM' component in the spin-flop phase.  Clearly, this complete correspondence does not always hold when the magnetic moment are along a low-symmetry direction.  An extreme case occur when the magnetic moments are along a completely generic direction $[x,y,z]$, since in this case the MPG analysis yields six independent components along $[x,y,z]$.  However, these situations are extremely rare, and can be usually resolved within the MPG framework by rotating the magnetic moment along a high-symmetry direction (but see here below for the case of cubic groups).

\subsection{Strong-collinear altermagnetism in cubic groups}
\label{sec: cubic_groups}

The relation between the SG and MPG approaches becomes even clearer when considering the case of the cubic groups.  Cubic MPGs do not admit collinear structures, since the direction of the magnetic moments always breaks the cubic symmetry.  However, cubic symmetry can still hold in an \emph{approximate} sense if the spin-lattice interaction is small.  Highlighting these approximate symmetries and their effects on momentum-space textures is one of the main advantages of SGs, which are therefore an extremely useful complement to MPGs.  As shown for other symmetries in sec. \ref{sec: ex_spingroups}, it is often possible to choose a high-symmetry spin direction where no space operator is lost, but this is clearly not possible for cubic groups.  It is therefore useful to work our a cubic example in detail, so as to understand exactly how the two approaches are related.

Let us consider AFM ordering on a crystal with point group $\hmn{m-3m}$, such as the pyrochlore Er$_2$Ru$_2$O$_7$ (ref. \onlinecite{Taira2003}), in which cubic symmetry is broken by a collinear AFM structure. The magnetic moments are along the $z$ axis, and the MPG is $\hmn{*4'm'm}$ (Class XI). The corresponding SG is $\hmn{m -3 \tilde{m}}$ (full symbol $\hmn{*{\tilde{4}} -3 !{\tilde{2}}}$)   \emph{regardless} of the direction of the magnetic moments, where, once again, SG space operators do not change the direction of the spins and the ``$\tilde{\;\;}$'' symbol indicates spin flip.  At the rank-7 level, the effective field components for the two cases are:

\begin{eqnarray}
\label{eq: cubic_fields}
 B^{SG} (\vec{k})&=&\Lambda(k)(k_x^2 - k_y^2) (k_x^2 - k_z^2) (k_y^2 - k_z^2)\nonumber\\
B_z^{MPG} (\vec{k})&=& B^{SG} (\vec{k})+\left(k_x^2 - k_y^2) (\Lambda_1(k) k_x^2 k_y^2 + \Lambda_2(k) (k_x^2 + k_y^2)^2+ \Lambda_3(k) k_z^4\right)
\end{eqnarray}

\begin{figure}[!h]
\centering
\includegraphics[scale=0.5, trim={0 3cm , 0, 3cm},clip]{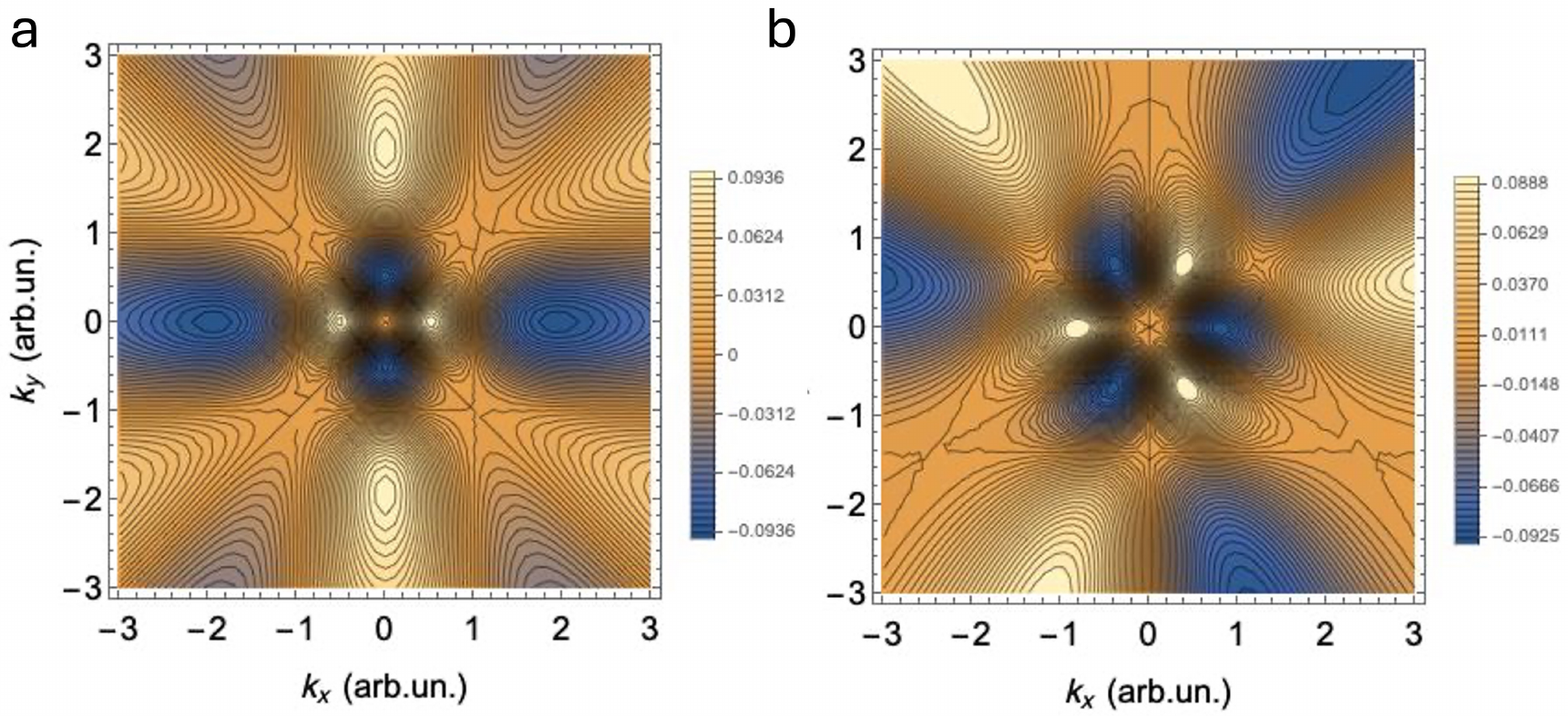}
\caption{ \label{fig: cubic}  Views along the $[001]$ (a) and $[111]$ (b) directions of the scalar texture pattern (i.e., the texture component along the direction of the magnetic moments) for MPG $\hmn{*4'm'm}$ (symmetrised to cubic) or spin group  $\hmn{*{\tilde{4}} -3 !{\tilde{2}}}$.  The normalisation function is as in previous figures.  Contours were plotted for $k_z=1$ and are both symmetric by exchange $k_z\rightarrow -k_z$.  Panel (a) to be compared with fig. 2, sub-panel B6 of ref. \onlinecite{Smejkal2022}.}
\end{figure}

Corresponding MPG expressions for magnetic moment in other high-symmetry directions and in the appropriate cooordinate systems are reported in tab. \ref{Tab: cubic_forms}.  For the tetragonal case, is clear that $B^{SG}(\vec{k})$ is the symmetrised version of $B_z^{MPG} (\vec{k})$, obtained by setting $\Lambda_1(k)=\Lambda_2(k)=\Lambda_3(k)=0$.  
The additional components would emerge as a consequence of symmetry breaking, and can be reasonably expected to be small for a collinear structure where cubic symmetry is only broken by the spin system and spin-lattice interaction is weak.  Fig. \ref{fig: cubic} shows the texture pattern for $B^{SG}(\vec{k})$ (i.e., in the absence of the tetragonal terms).   Fig. \ref{fig: cubic} also shows that the scalar texture pattern is \emph{completely symmetric} by the SG cubic symmetry operators, while the generic tetragonal, trigonal or orthorhombic textures for the corresponding MPGs would only have those specific symmetries.


\begin{figure}[!h]
\centering
\includegraphics[scale=0.5, trim={0 3cm , 0, 3cm},clip]{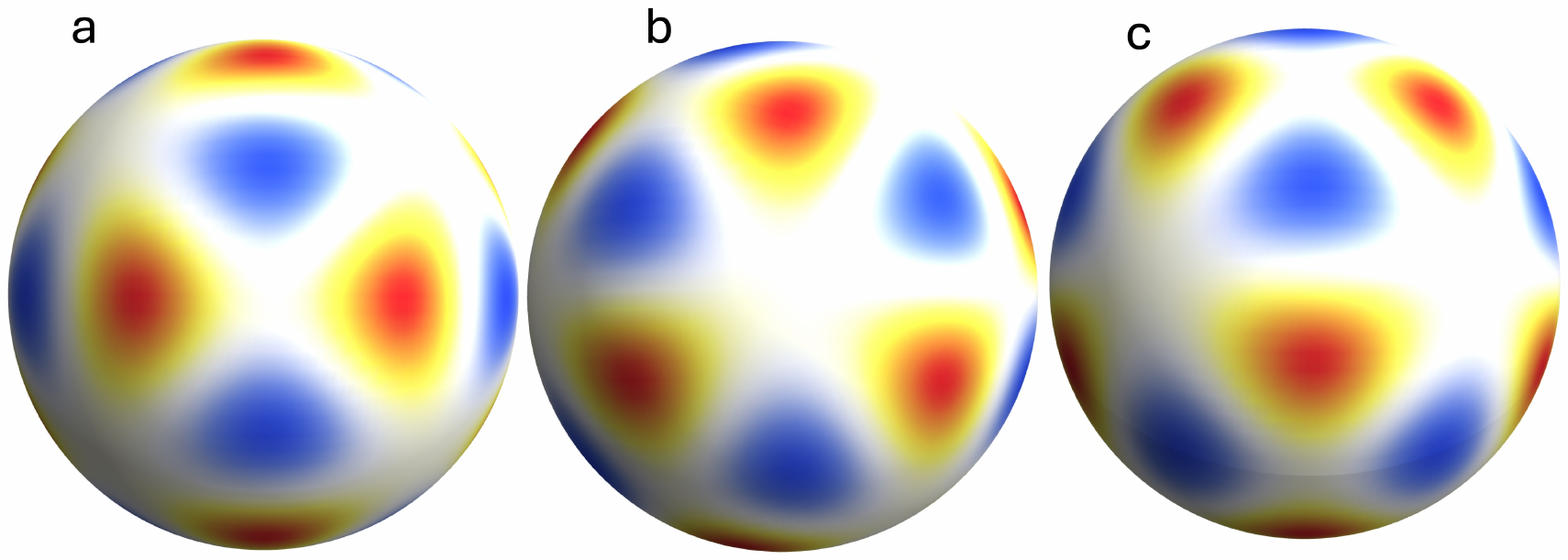}
\caption{ \label{fig: spheres}  Views along the $[001]$ (a) and $[111]$ (b) and $[110]$ (c) directions of the scalar spin texture in fig. \ref{fig: cubic} plotted on the surface of a sphere in momentum space.  Blue/red colours correspond to negative/positive values.}
\end{figure}

For further clarity, views along different directions of these textures plotted on the surface of a sphere in momentum space are displayed in  fig. \ref{fig: spheres}.  As remarked earlier, textures for different values of $k=|\vec{k}|$ only differ by the multiplicative function $\Lambda(k)$ (see first line of eq. \ref{eq: cubic_fields}).

\begin{tiny}
\begin{table*}[h!]
\centering
\caption{\label{Tab: cubic_forms} Polynomial forms of the effective field components for cubic groups and different high-symmetry magnetic moment directions.  The corresponding spin groups are indicated in the first column.  For each moment direction, the corresponding MPG class is indicated together with one MPG example (to establish the correct setting). The symbol $\hmn{..2}$ means that the 2-fold axis is along the $z$ direction (see text).}  
\begin{tabular}{ |c|c|c|c|c| } 
\hline
Spin groups &  & $S \parallel [001]$ & $S \parallel [110]$ & $S \parallel [111]$\\
\hline
\multirow{2}{4em}{$\hmn{23}$,$\hmn{m-3}$ } & Class & II ($\hmn{..2}$)& I ($\hmn{1}$)&VIII ($\hmn{3}$)\\ \cline{2-5}
& Form &  $\Lambda(k) \left( k_x^2+k_y^2+k_z^2\right)$&$\Lambda(k) \left (k_x^2+k_y^2+k_z^2\right)$ &$\Lambda(k) \left (k_x^2+k_y^2+k_z^2\right)$\\
\hline
$\hmn{432}$, $\hmn{-43m}$& Class & XIVb ($\hmn{42'2'}$)& IV ($\hmn{2'2'2}$)&VII ($\hmn{32'}$)\\ \cline{2-5}
$\hmn{m-3m}$& Form &  $\Lambda(k) \left( k_x^2+k_y^2+k_z^2\right)$&$\Lambda(k) \left (k_x^2+k_y^2+k_z^2\right)$ &$\Lambda(k) \left (k_x^2+k_y^2+k_z^2\right)$\\
\hline
$\hmn{\tilde{4}3\tilde{2}}$& Class & XI ($\hmn{*4'm'm}$)& IV ($\hmn{2'2'2}$)&VIII ($\hmn{32}$)\\ \cline{2-5}
$\hmn{\tilde{{-4}}3\tilde{m}}$,$\hmn{m -3 \tilde{m}}$& Form& $\begin{array} {c} \Lambda(k)(k_x^2 - k_y^2) \cdot \\(k_x^2 - k_z^2) (k_y^2 - k_z^2)\end{array}$ & $\begin{array} {c} \Lambda(k) (-2 k_x^2 + (k_y - k_z)^2)\cdot \\  \left(-2 k_x^2 + (k_y + k_z)^2\right)k_y k_z\end{array}  $&$\begin{array}{c} \Lambda(k)  (k_x^2 - 3 k_y^2)k_x k_z \cdot \\ \left [ 6(k_x^2 + k_y^2) - 16 k_z^2 - \sqrt{2} ( k_y^2 - 3 k_x^2) \right]\end{array} $\\
\hline
\end{tabular}
\end{table*}
\end{tiny}

\section{Discussion}
\label{sec: discussion}

In this paper, I have outlined a natural symmetry classification of all $\vec{k}/-\vec{k}$-symmetric, time-reversal-odd altermagnets, based on the lowest-rank altermagnetic tensor forms allowed in each magnetic point group (MPG).  I have also tabulated all the higher-rank tensor forms required to describe `strong-collinear' altermagnetic textures, i.e., those that produce spin splitting with quantisation axis parallel to the N\'eel vector.  This include the case of cubic groups, where the spin-group formalism (thoroughly developed by other groups)\cite{Smejkal2022} is most useful. 

This classification has important practical consequences, particularly when one considers the tensorial connection between altermagnetism (which describes particular form of reciprocal-space spin textures) and well-known macroscopic properties such as piezomagnetism and the MOKE effect.  For example, the Bilbao database MAGNDATA \cite{Gallego2016} lists 139 known materials that are MOKE active and do not allow ferromagnetism.  The point groups of these materials are the same as those in classes V, VIII, IX, X, XI, XII, XIV, XV and XVII in which purely antiferromagnetic quadratic altermagnetism is allowed.  In fact, most materials so far discussed in the context of altermagnetims (and many more) are well-known piezomagnets/MOKE-active AFM.  Moreover, while a positive identification of altermagnetism as distinct from time-even effects such as the surface R-D effect is rather difficult, detecting MOKE activity is rather straightforward and would enable screening of candidate materials for which the magnetic structural determination may be ambiguous. \cite{Kessler2024}

One may also observe that altermagnetism is allowed in 69 MPGs, of which 66 allow quadratic altermagnetism and 31 allow ferromagnetism.  Considering that 32 MPGs are paramagnetic (grey), out of a total of 122 MPGs only 21 groups are black\&white and do not allow altermagnetism, which is therefore hardly a rare phenomenon.

Although the purpose of this paper is mainly to discuss altermagnetic \emph{symmetry}, a few words about `mechanisms' do not seem inappropriate, particularly in connection with the role of the spin-orbit coupling (SOC).  As already discussed in ref. \onlinecite{Yuan2020}, altermagnetic splitting is caused by the time-reversal-odd effective Zeeman fields (mostly of exchange origin) generated by real magnetic moments in the unit cell .  Since the SO interaction is time-reversal even (its sources being the gradients of the electrical potential), at first sight it would seem that the SO interaction could not be responsible for $\vec{k}/-\vec{k}$-symmetric, time-reversal-odd altermagnetism.

However, this in not entirely correct, since the SOC can affect both the orientation of the localised magnetic moments and the spin density distribution, most famously by inducing spin canting via the Dzyaloshinskii-Moriya interaction.  It is therefore likely that the SOC is responsible for `weak altermagnetism', \cite{Krempasky2024} particularly when spin splitting along a unique high-symmetry direction is not allowed at the lowest order (Classes VIII, XII, XIV, XV, XVII and XVIII).  Interestingly, one must also expect to observe R-D-like, $\vec{k}/-\vec{k}$-\emph{anti}symmetric splitting that is of pure magnetic origin, because magnetism ordering can break inversion symmetry.   The two best-known cases of this phenomenon, described by Cheong \textit{et al.} as `type-III altermagnetism'\cite{Cheong2024} and discussed theoretically in two recent publications, \cite{Hayami2022, Hellenes2023} are helical magnetic structures and type-II multiferroicity, both of which can originate in centrosymmetric crystal due to competing interactions.\footnote{Altermagnetism as defined in this paper is not allowed in paramagnetic (grey) point groups. However, many antiferromagnets (especially incommensurate AFMs) possess paramagnetic MPGs, because in there MSG time reversal is equivalent to a (potentially incommensurate) translation U --- in other words, $U\theta$ is a symmetry operator.  This is the case for most (but not all) non-$\Gamma$-point AFMs.  If, in addition, inversion is broken, these systems are candidates for `Type III altermagnetism'.  A systematic treatment of these system similar to the one presented here should be possible since, even in the most complex cases, one can always define a MPG.  \cite{Radaelli2007}}  In both cases, the relevant properties (magnetic helicity and magnetic polarity, respectively) are \emph{time reversal even}, so this materials do not violate the general rule that the spin splitting should be the same in time-reversed domains.





\appendix

\section{Tensor equalities and invariance}
\subsection{Equivalence of Cartesian and Spherical tensor decompositions}

\label{App_A1}

In general, Cartesian tensors like those in eq. \ref{eq. tensor_expansion} are members of the tensor product space $\Gamma^V \otimes \left[\Gamma^V \otimes  \Gamma^V \dots \right]$, where the square bracket indicates index symmetrisation while $\Gamma^V$ is the representation of ordinary vectors, which coincides with the $L=1$ \textit{irrep} of $SO(3)$ ($\Gamma^1$) if one considers only proper rotations.   In turn, $\left[\Gamma^V \otimes  \Gamma^V \dots \right]$ can be decomposed in $SO(3)$ \textit{irreps} as follows:


\begin{eqnarray}
\left[\Gamma^V \otimes  \Gamma^V  \right]&=& \Gamma^2+\Gamma^0\nonumber\\
\left[\Gamma^V \otimes  \Gamma^V\otimes  \Gamma^V  \right]&=&\Gamma^3+\Gamma^1\nonumber\\
\left[\Gamma^V \otimes  \Gamma^V\otimes  \Gamma^V\otimes  \Gamma^V  \right]&=&\Gamma^4+\Gamma^2+\Gamma^0\nonumber\\
\left[\Gamma^V \otimes  \Gamma^V\otimes  \Gamma^V\otimes  \Gamma^V\otimes  \Gamma^V  \right]&=&\Gamma^5+\Gamma^3+\Gamma^1\nonumber\\
\dots&&
\end{eqnarray}

where $\Gamma^2, \Gamma^3$, etc. are the \textit{irreps} for $L=2, 3, \dots$. Taking once again the tensor product with  $\Gamma^V$

\begin{eqnarray}
\Gamma^V \otimes\left[(\Gamma^V) ^2  \right]&=& \Gamma^3+\Gamma^2+2 \Gamma^1\nonumber\\
\Gamma^V \otimes\left[(\Gamma^V) ^3 \right]&=&\Gamma^4+\Gamma^3+2 \Gamma^2+\Gamma^1+\Gamma^0\nonumber\\
\Gamma^V \otimes\left[(\Gamma^V) ^4  \right]&=&\Gamma^5+\Gamma^4+2 \Gamma^3 +\Gamma^2+2 \Gamma^1\nonumber\\
\Gamma^V \otimes\left[(\Gamma^V) ^5  \right]&=&\Gamma^6+\Gamma^5+2 \Gamma^4+\Gamma^3+2 \Gamma^2+\Gamma^1+\Gamma^0\nonumber\\
\dots
\end{eqnarray}

where I have used the shorthand $(\Gamma^V) ^2=\Gamma^V \otimes \Gamma^V$ etc.  The spherical tensor decomposition is the familiar one:

\begin{eqnarray}
\label{eq: spharms}
\Gamma^1 \otimes \Gamma^0 &=& \Gamma^1\nonumber\\
\Gamma^1 \otimes \Gamma^1 &=& \Gamma^2+\Gamma^1+ \Gamma^0\nonumber\\
\Gamma^1 \otimes \Gamma^2&=&\Gamma^3+\Gamma^2+\Gamma^1\nonumber\\
\Gamma^1 \otimes \Gamma^3&=&\Gamma^4+\Gamma^3+\Gamma^2\nonumber\\
\Gamma^1 \otimes \Gamma^4&=&\Gamma^5+\Gamma^4+\Gamma^3\nonumber\\
\Gamma^1 \otimes \Gamma^5&=&\Gamma^6+\Gamma^5+\Gamma^4\nonumber\\
\dots
\end{eqnarray}

So, for example, the sum of all the spherical harmonics terms in eq. \ref{eq: spharms} is:

\begin{eqnarray}
\sum_{n=0}^5 \Gamma^1 \otimes \Gamma^n&=&\Gamma^0+3 \Gamma^1+3\Gamma^2+3\Gamma^3+3 \Gamma^4 + 2 \Gamma^5 + \Gamma^6\nonumber\\
&=& \Gamma^V \otimes\left[(\Gamma^V) ^4\right] + \Gamma^V \otimes\left[(\Gamma^V) ^5\right]
\end{eqnarray}

We can therefore conclude that, if one considers only proper rotations (i.e., elements of the continuous group $SO(3)$), the decomposition into spherical tensors of a vector field on a spherical shell up to order $n$ is equivalent to the sum of two Cartesian tensors of ranks $n$ and $n+1$.  Including parity and time reversal, it is found that the Cartesian tensors of odd rank are parity-even and time-reversal odd, while the Cartesian tensors of even rank are parity-odd and time-reversal even.

\subsection{Tensor invariance}
\label{App_A2}

 Here, I show that the symmetry of the global point-group symmetry of spin texture generated by a certain Cartesian tensor requires that tensor to be totally symmetric by the symmetry operations of that point group.  In general, given a vector field $\vec{V}(\vec{k})$ and a proper rotation $\mat{R}$ the transformed vector field is:

 \begin{equation}
 \label{eq: transf1}
 \tilde{\vec{V}}(\vec{k})= \mat{R} \cdot \vec{V}(\mat{R}^{-1} \cdot \vec{k})
 \end{equation}
 
 Particularising to the case of $\vec{B}^{eff}(\vec{k})$ we obtain for \emph{proper} rotations:
 
 \begin{eqnarray}
 \label{eq: transf2}
\tilde{\vec{B}}^{eff}(\vec{k})&=&R_{ij} T^{(n)}_{j,\alpha\beta\gamma\dots} R^{-1}_{\alpha \rho} R^{-1}_{\beta \sigma} R^{-1}_{\gamma \tau} k_\rho k_\sigma k_\tau \dots \nonumber\\
&=& R_{ij} R_{\rho \alpha} R_{ \sigma \beta} R_{\tau \gamma}  T^{(n)}_{j,\alpha\beta\gamma\dots}  k_\rho k_\sigma k_\tau\dots\nonumber\\
\end{eqnarray}

 After some bookeeping can generalise this to proper/improper rotations that may include time reversal by considering that $\vec{k}$ is time-reversal-odd and parity-odd, as:
 
\begin{eqnarray}
 \label{eq: transf3}
\tilde{\vec{B}}^{eff}(\vec{k})&=& (-1)^{(n+1) p +n t}\nonumber\\
&& R_{ij} R_{\rho \alpha} R_{ \sigma \beta} R_{\tau \gamma} \dots T^{(n)}_{j,\alpha\beta\gamma\dots}  k_\rho k_\sigma k_\tau\dots\nonumber\\
\end{eqnarray}

If we now require that, $\forall \vec{k}$, $\tilde{\vec{B}}^{eff}(\vec{k})=\vec{B}^{eff}(\vec{k})$ , we conclude:

\begin{eqnarray}
 \label{eq: transft}
(-1)^{(n+1) p +n t} R_{ij} R_{\rho \alpha} R_{ \sigma \beta} R_{\tau \gamma} \dots T^{(n)}_{j,\alpha\beta\gamma\dots} &=&  T^{(n)}_{i,\rho\sigma\tau\dots}\nonumber\\
\end{eqnarray}

which is precisely the definition of a tensor that is totally symmetric by a generalised rotation $\mat{R}$ that may include inversion and time reversal 

\section{Spherical basis tensors}
\label{App_B}

In Table \ref{Table_2}, I present a decomposition of the 17 unique tensor forms for quadratic altermagnetism into a set of nine simple spherical basis tensors plus those obtained by axis permutations.  Each of the 17 tensor forms can be obtained as a linear combination of the spherical basis tensors (Table \ref{Table_3}).  For example, for Class V, we have:

\begin{widetext}
\begin{eqnarray}
\left(
\begin{array}{cccccc}
 0 & 0 & 0 & \Lambda _{14} & 0 & 0 \\
 0 & 0 & 0 & 0 & \Lambda _{25} & 0 \\
 0 & 0 & 0 & 0 & 0 & \Lambda _{36} \\
\end{array}
\right) &=&  \Lambda_1 Q^{I\!I} +  \Lambda_2 Q^{I\!I\!I} +  \Lambda_3 O^{I} =\nonumber\\
&=& \Lambda_1 \left(
\begin{array}{cccccc}
 0 & 0 & 0 & -1 & 0 & 0 \\
 0 & 0 & 0 & 0 & 1 & 0 \\
 0 & 0 & 0 & 0 & 0 & 0 \\
\end{array}
\right)+ \Lambda_2 \left(
\begin{array}{cccccc}
 0 & 0 & 0 & 1 & 0 & 0 \\
 0 & 0 & 0 & 0 & 1 & 0 \\
 0 & 0 & 0 & 0 & 0 & -2 \\
\end{array}
\right) +\Lambda_3 \left(
\begin{array}{cccccc}
 0 & 0 & 0 & 1 & 0 & 0 \\
 0 & 0 & 0 & 0 & 1 & 0 \\
 0 & 0 & 0 & 0 & 0 & 1 \\
\end{array}
\right)\nonumber\\
\end{eqnarray}  
\end{widetext}

with

\begin{eqnarray}
\Lambda _{14}&=&-\Lambda_1+\Lambda_2+\Lambda_3\nonumber\\
\Lambda _{25}&=&\Lambda_1+\Lambda_2+\Lambda_3\nonumber\\
\Lambda _{36}&=&-2 \Lambda_2+\Lambda_3\nonumber\\
\end{eqnarray}


\section{Binary spin groups in the time-reversal language}
\label{App: spin_groups}
Spin groups (SG) are generally defined as subgroups of the  outer direct product of a point (or space) group and a spin rotation group. \cite{Litvin1977} The simplest non-trivial SGs are \emph{binary} SG, in which the group acting on spins is a 2-element set. To classify collinear structures \v{S}mejkal \textit{et al.} adopt the two-element group $\{1, 2_z\}$, where $1$ is the identity and $2_z$ is a 2-fold rotation perpendicular to the N\'eel vector.  Another choice is to use the two-element group $\{1, 1'\}$, where $1'$ is the time reversal operator. With either choice, binary SG coincide with the the celebrated Shubnikov groups. \cite{Opechowski1965}  The only difference between binary SG with the time-reversal operator and the MPG treatment (which also employs Shubnikov point groups) is that binary SG are made to act on \emph{scalar, time-reversal-odd textures} both in real and in reciprocal space (space operators have no effect on scalars), while MPGs act on \emph{axial-vector, time-reversal-odd textures}.  For binary SGs, I have indicated the time-reversal operator with the symbol ``$\tilde{\;\;}$'' to distinguish it from the MPG equivalent, indicated with a prime (') This choice enables one to recast the action of binary SGs onto the familiar language of parity, time reversal etc., and to derive a parallel tensorial treatment to that of MPGs.  

For a given collinear structure, converting SG into MPGs and vice versa requires at the very least knowledge of the direction of the N\'eel vector. However, in the presence of pseudo-symmetry (i.e., when crystal symmetry is broken only by the direction of the magnetic moments), there is no simple correspondence between its MPGs and binary SG, with the latter generally having more symmetry operators than the former. For example, for a tetragonal structure with binary SG $\hmn{*{\tilde{4}}\tilde{m}m}$, the MPG is $\hmn{*{4'}mm'}$  for magnetic moments along the $z$ axis and $\hmn{m'mm'}$ or $\hmn{mm'm'}$ for N\'eel vector along $x$ and $y$, respectively.  Several examples of these conversions are given in sec. \ref{sec: examples}.

\section{Construction of gnomonic projections from band dispersions}
\label{app: gnomonic}

Here, I explain how to construct gnomonic projections of spin textures, similar to those displayed in the figures of sec. \ref{sec: examples}, starting from spin-polarised band dispersions such as those obtained from DFT calculations.  One starts by calculating a grid of points at constant $k=|\vec{k}|$ for a particular band.  It is most convenient to calculate points at constant $\theta$ and equal intervals $\Delta \phi$, where $k$, $\theta$ and $\phi$ are spherical coordinates in momentum space.  To reproduce the figures of sec. \ref{sec: examples}, one plots the spin polarisation for points at constant $\theta$ on a circle of radius $\tan \theta$ and with the original values of $\phi$.  Figures constructed from points calculated by DFT will include the pre-factors $\Lambda_i(k)$, which depend on both $k$ and the band index.  Moreover, one may require higher-rank tensors than those discussed in the paper to reproduce DFT data accurately.

\acknowledgments{I acknowledge discussions with A. Stroppa (CNR-SPIN), Roger D. Johnson (University College London), Dmitry Khalyavin (STFC UKRI)  and Gautam Gurung (Trinity College, Oxford).



\afterpage{
\begin{widetext} 
\begin{tiny}
\tablecaption{\label{Table_2} Symmetry-adapted spherical basis tensors employed for the decomposition of the altermagnetic tensor in each of the 17 classes (see Table \ref{Table_1}).  The basis tensors are constructed from  9 unique forms (two dipolar, three quadrupolar and four octupolar) and their axis permutations.  Only permutations actually employed in the decomposition are listed.}

\begin{xtabular*}{\textwidth}{l@{\extracolsep{\fill}}llll}

\hline\hline\hline
\multicolumn{4}{c}{\textbf{}}\\
\textbf{Spherical form} & \textbf{x} & \textbf{y} &\textbf{ z}\\
\\
\hline
\\
D$^{I}$&$\left(
\begin{array}{rrrrrr}
 $1$ & $1$ & $1$ & $0$ & $0$ & $0$ \\
 $0$ & $0$ & $0$ & $0$ & $0$ & $0$ \\
 $0$ & $0$ & $0$ & $0$ & $0$ & $0$ \\
\end{array}
\right)$&
$
\left(
\begin{array}{rrrrrr}
 $0$ & $0$ & $0$ & $0$ & $0$ & $0$ \\
$1$ & $1$ & $1$ & $0$ & $0$ & $0$ \\
 $0$ & $0$ & $0$ & $0$ & $0$ & $0$ \\
\end{array}
\right)
$
&
$
\left(
\begin{array}{rrrrrr}
 $0$ & $0$ & $0$ & $0$ & $0$ & $0$ \\
 $0$ & $0$ & $0$ & $0$ & $0$ & $0$ \\
 $1$ & $1$ & $1$ & $0$ & $0$ & $0$ \\
\end{array}
\right)
$\\
\\
D$^{I\!I}$&
$
\left(
\begin{array}{rrrrrr}
 $1$ & $0$ & $0$ & $0$ & $0$ & $0$ \\
 $0$ & $0$ & $0$ & $0$ & $0$ & $1$ \\
 $0$ & $0$ & $0$ & $0$ & $1$ & 0 \\
\end{array}
\right)
$
&
$
\left(
\begin{array}{rrrrrr}
 $0$ & $0$ & $0$ & $0$ & $0$ & $1$ \\
 $0$ & $1$ & $0$ & $0$ & $0$ & $0$ \\
 $0$ & $0$ & $0$ & $1$ & $0$ & $0$ \\
\end{array}
\right)
$
&
$
\left(
\begin{array}{rrrrrr}
 $0$ & $0$ & $0$ & $0$ & $1$ & $0$ \\
 $0$ & $0$ & $0$ & $1$ & $0$ & $0$ \\
 $0$ & $0$ & $1$ & $0$ & $0$ & $0$ \\
\end{array}
\right)
$
\\
\\
\hline
\\

Q$^{I}$&
$
\left(
\begin{array}{rrrrrr}
 $0$ & $1$ & $-1$ & $0$ & $0$ & $0$ \\
 $0$ & $0$ & $0$ & $0$ & $0$ & $-1$ \\
 $0$ & $0$ & $0$ & $0$ & $1$ & $0$ \\
\end{array}
\right)
$
&
$
\left(
\begin{array}{rrrrrr}
 $0$ & $0$ & $0$ & $0$ & $0$ & $1$ \\
$ -1$ & $0$ & $1$ & $0$ & $0$ & $0$ \\
 $0$ & $0$ & $0$ & $-1$ & $0$ & $0$ \\
\end{array}
\right)
$
\\
\\
Q$^{I\!I}$ &&&
$
\left(
\begin{array}{rrrrrr}
 $0$ & $0$ & $0$ & $-1$ & $0$ & $0$ \\
 $0$ & $0$ & $0$ & $0$ & $1$ & $0$ \\
 $0$ & $0$ & $0$ & $0$ & $0$ & $0$ \\
\end{array}
\right)
$
\\
\\
Q$^{I\!I\!I}$&&&
$
\left(
\begin{array}{rrrrrr}
 $0$ & $0$ & $0$ & $1$ & $0$ & $0$ \\
 $0$ & $0$ & $0$ & $0$ & $1$ & $0$ \\
 $0$ & $0$ & $0$ & $0$ & $0$ & $-2$ \\
\end{array}
\right)
$
\\
\\
\hline
\\
O$^{I}$&
$
\left(
\begin{array}{rrrrrr}
 $0$ & $0$ & $0$ & $1$ & $0$ & $0$ \\
 $0$ & $0$ & $0$ & $0$ & $1$ & $0$ \\
 $0$ & $0$ & $0$ & $0$ & $0$ & $1$ \\
\end{array}
\right)
$
\\
\\
O$^{I\!I}$&
&
&
$
\left(
\begin{array}{rrrrrr}
 $0$ & $0$ & $0$ & $0$ & $2$ & $0$ \\
$0$ & $0$ & $0$ & $-2$ & $0$ & $0$ \\
 $1$ & $-1$ & $0$ & $0$ & $0$ & $0$ \\
\end{array}
\right)
$
\\
\\
\\
O$^{I\!I\!I}$&
$
\left(
\begin{array}{rrrrrr}
 $-1$ & $1$ & $0$ & $0$ & $0$ & $0$ \\
 $0$ & $0$ & $0$ & $0$ & $0$ & $2$ \\
$0$ & $0$ & $0$ & $0$ & $0$ & $0$ \\
\end{array}
\right)
$
&
$
\left(
\begin{array}{rrrrrr}
 $0$ & $0$ & $0$ & $0$ & $0$ & $2$ \\
 $1$ & $-1$ & $0$ & $0$ & $0$ & $0$ \\
 $0$ & $0$ & $0$ & $0$ & $0$ & $0$ \\
\end{array}
\right)
$
&\\
\\
O$^{I\!V}$&
$
\left(
\begin{array}{rrrrrr}
 $-2$ & $1$ & $1$ & $0$ & $0$ & $0$ \\
 $0$ & $0$ & $0$ & $0$ & $0$ & $2$ \\
 $0$ & $0$ & $0$ & $0$ & $2$ & $0$ \\
\end{array}
\right)
$
&
$
\left(
\begin{array}{rrrrrr}
 $0$ & $0$ & $0$ & $0$ & $0$ & $2$ \\
 $1$ & $-2$ & $1$ & $0$ & $0$ & $0$ \\
 $0$ & $0$ & $0$ & $2$ & $0$ & $0$ \\
\end{array}
\right)
$ &
$
\left(
\begin{array}{rrrrrr}
 $0$ & $0$ & $0$ & $0$ & $2$ & $0$ \\
 $0$ & $0$ & $0$ & $2$ & $0$ & $0$ \\
 $1$ & $1$ & $-2$ & $0$ & $0$ & $0$ \\
\end{array}
\right)
$
\\
\\

\hline\hline\hline

\end{xtabular*}
\end{tiny}
\end{widetext} 
}


\newpage

\afterpage{
\begin{widetext} 
\begin{tiny}
\tablecaption{\label{Table_3} Decomposition of the tensors for the 17 altermagnetic classes in terms of the symmetry-adapted spherical basis tensors in Table \ref{Table_2}.  The subscripts indicates the appropriate axis permutation (see columns in Table \ref{Table_2}) and is omitted when it is not ambiguous.}

\begin{xtabular*}{\textwidth}{l@{\extracolsep{\fill}}lllll}

\hline\hline\hline
\multicolumn{3}{c}{\textbf{}}\\
\textbf{Class} & \textbf{Magnetic point groups} & \textbf{Spherical basis}\\
\\
\hline
\\
\textbf{Class I}$^\S$ &
\begin{tabular}{cc}
\hmn{1} & \hmn{-1}
\end{tabular} 
& D$^{I}_x$ + D$^{I}_y$ + D$^{I}_z$ + D$^{I\!I}_x$ + D$^{I\!I}_y$ + D$^{I\!I}_z$ + Q$^{I}_x$ + 
 Q$^{I}_y$ + Q$^{I}_z$ + Q$^{I\!I}$ +  \\
 
 &&Q$^{I\!I\!I}$ + O$^{I}$ + O$^{I\!I}_z$ + O$^{I\!I\!I}_x$ +O$^{I\!I\!I}_y$ + O$^{I\!V}_x$ + O$^{I\!V}_y$ + O$^{I\!V}_z$\\
\\ 
\textbf{Class II}$^\S$ & 
\begin{tabular}{ccc}
\hmn{2}&\hmn{m}& \hmn{*2} 
\end{tabular} 
& D$^{I}_y$ + D$^{I\!I}_y$ + Q$^{I}_y$ + Q$^{I\!I}$ + Q$^{I\!I\!I}$ + O$^{I}$ + O$^{I\!I\!I}_y$ + O$^{I\!V}_y$  \\
\\
\textbf{Class III}$^\S$ & 
\begin{tabular}{ccc}
\hmn{2'}& \hmn{m'}& \hmn{`2} 
\end{tabular} 
& D$^{I}_x$ + D$^{I}_z$ + D$^{I\!I}_x$ + D$^{I\!I}_z$ + Q$^{I}_x$ + Q$^{I}_z$ + O$^{I\!I\!I}_x$ + 
 O$^{I\!V}_x$ + O$^{I\!I}_z$ + O$^{I\!V}_z$  \\
\\
\textbf{Class IV}$^\S$ & 
\renewcommand{\arraystretch}{1.5}
\begin{tabular}{cc}
\hmn{2'2'2}&\hmn{m'm2'}\\
\hmn{m'm'2}&\hmn{m'm'm}
\end{tabular}  & D$^{I}_z$ + D$^{I\!I}_z$ + Q$^{I}_z$ + O$^{I\!I}_z$, + O$^{I\!V}_z$ \\
\\
\\
Class V$^\dagger$ & 
\begin{tabular}{ccc}
\hmn{222}& \hmn{mmm}& \hmn{mm2}
\end{tabular} 
 & Q$^{I\!I}$ + Q$^{I\!I\!I}$ + O$^{I}$  \\
\\
\textbf{Class VI}$^\S$ & 
\begin{tabular}{cc}
\hmn{3}&\hmn{-3}
\end{tabular} 
& D$^{I}_z$ + D$^{I\!I}_z$ + Q$^{I\!I}$ + O$^{I\!I\!I}_x$ + O$^{I\!I\!I}_y$ + O$^{I\!V}_z$  \\
\\
\textbf{Class VII}$^\S$ & 
\begin{tabular}{ccc}
\hmn{32'}&\hmn{3m'}&\hmn{-3m'}
\end{tabular} 
& D$^{I}_z$ + D$^{I\!I}_z$ + O$^{I\!I\!I}_y$ + O$^{I\!V}_z$  \\
\\
\textit{Class VIII}$^*$& \begin{tabular}{ccc}
\hmn{32}&\hmn{3m}&\hmn{-3m}
\end{tabular}
&  Q$^{I\!I}$+O$^{I\!I\!I}_x$  \\
\\
Class IX$^\dagger$ & \begin{tabular}{ccc}
\hmn{4'}&\hmn{-4'}&\hmn{*{4'}}
\end{tabular}
& Q$^{I}_z$ + Q$^{I\!I\!I}$ + O$^{I}$ + O$^{I\!I}_z$  \\
\\
Class X$^\dagger$ &\begin{tabular}{cc}
\hmn{4'22'}&\hmn{-4'2m'}
\end{tabular}
& Q$^{I\!I\!I}$ + O$^{I}$ \\
\\
Class XI$^\dagger$ & 
\begin{tabular}{ccc}
\hmn{-4'2'm}&\hmn{4'm'm}&\hmn{*{4'}m'm}
\end{tabular}
& Q$^{I}_z$ + O$^{I\!I}_z$  \\
\\
\textit{Class XII}$^*$ & 
\begin{tabular}{ccc}
\hmn{6'}&\hmn{-6'} & \hmn{`{6'}}
\end{tabular}
&O$^{I\!I\!I}_x$ + O$^{I\!I\!I}_y$  \\
\\
\textbf{Class XIII}$^\S$ & 
\renewcommand{\arraystretch}{1.5}
\begin{tabular}{cccc}
(a)&\hmn{6}&\hmn{-6}&\hmn{*6}\\
(b)&\hmn{4}&\hmn{-4}&\hmn{*4} 
\end{tabular}
& D$^{I}_z$ + D$^{I\!I}_z$ + Q$^{I\!I}$ + O$^{I\!V}_z$  \\
\\
\textit{Class XIV}$^*$ & 
\renewcommand{\arraystretch}{1.5}
\begin{tabular}{ccc}
\hmn{6'22'}&\hmn{-6'2m'}&\hmn{-6'm2'}\\
\hmn{6'mm'}&\hmn{`{6'}mm'}
\end{tabular} 
 & O$^{I\!I\!I}_x$  \\
\\
\textit{Class XV}$^*$ & 
\renewcommand{\arraystretch}{1.5}
\begin{tabular}{ccccc}
(a)&\hmn{622}& \hmn{-6m2} & \hmn{6mm}& \hmn{*6mm}\\
(b)& \hmn{422}&\hmn{-42m}&\hmn{4mm}&\hmn{*4mm}
\end{tabular} 
& Q$^{I\!I}$  \\
\\
\textbf{Class XVI}$^\S$ & 
\renewcommand{\arraystretch}{1.5}
\begin{tabular}{ccccc}
(a)&\hmn{62'2'}&\hmn{-6m'2'}&\hmn{6m'm'}&\hmn{*6m'm'}\\
(b)&\hmn{42'2'}&\hmn{-42'm'}&\hmn{4m'm'}&\hmn{*4m'm'}
\end{tabular} 
& D$^{I}_z$+D$^{I\!I}_z$+O$^{I\!V}_z$  \\
\\
\textit{Class XVII}$^*$ & \renewcommand{\arraystretch}{1.5}
\begin{tabular}{cccc}
(a)& \hmn{4'32'}&\hmn{-4'3m'}&\hmn{m-3m'} \\
(b)&\hmn{23}&\hmn{m-3}
\end{tabular}
& O$^{I}$  \\

\hline\hline\hline

\end{xtabular*}
\begin{tabular}{l}
(\S) Classes allowing a magnetic dipole term (Type-I altermagnets in ref. \onlinecite{Cheong2024})\\
($\dagger$) Classes allowing spin splitting along a permitted collinear AFM direction (`strong-collinear' altermagnetic classes)\\
(*) Classes not allowing spin splitting along a permitted collinear AFM direction (`weak-collinear' altermagnetic classes)\\
\end{tabular}
\end{tiny}
\end{widetext} 

}

%

\clearpage
\newpage

\bibliography{Altermag_Tensor_2024}

\end{document}